\documentclass[manuscript]{acmart}
\AtBeginDocument{%
  }

\setcopyright{acmlicensed}
\copyrightyear{2026}
\acmYear{2026}
\acmDOI{XXXXXXX.XXXXXXX}
\acmConference[CHI 2026]{ACM CHI conference on Human Factors in Computing Systems}{April 13--17,
  2026}{Barcelona, Spain}
\acmISBN{978-1-4503-XXXX-X/2018/06}

\usepackage{subcaption}
\usepackage{tabularx}

\usepackage{amsmath}       

\usepackage{amssymb}       
\usepackage{amsthm}        
\usepackage{mathtools}     
\usepackage{bm}            
\usepackage{bbm}           

\usepackage{tabularx}
\usepackage{tcolorbox}
\usepackage{booktabs} 

\usepackage{framed}
\usepackage{enumitem}

\usepackage{longtable} 
\usepackage{array}     

\usepackage{verbatim}

\usepackage[suppress]{color-edits}
\addauthor[Rebuttal]{re}{blue}
\addauthor[Rebuttal]{na}{blue}

\usepackage{xcolor}

\begin{document}

\title{LLM or Human? Perceptions of Trust and Information Quality in Research Summaries}


\author{Nil-Jana Akpinar}
\email{niljana.akpinar@gmail.com}
\affiliation{%
  \institution{Microsoft}
  \country{USA}
}
\authornote{This work was completed while the author was at Amazon AWS AI.}

\author{Sandeep Avula}
\email{sandeavu@amazon.com}
\affiliation{%
  \institution{Amazon AWS AI}
  \country{USA}
}

\author{CJ Lee}
\email{cjlee@amazon.com}
\affiliation{%
  \institution{Amazon AWS AI}
  \country{USA}
}

\author{Brandon Dang}
\email{dangbran@amazon.com}
\affiliation{%
  \institution{Amazon AWS AI}
  \country{USA}
}

\author{Kaza Razat}
\email{razat@amazon.com}
\affiliation{%
  \institution{Amazon AWS AI}
  \country{USA}
}

\author{Vanessa Murdock}
\email{vmurdock@acm.org}
\affiliation{%
  \institution{Amazon AWS AI}
  \country{USA}
}

\renewcommand{\shortauthors}{Akpinar et al.}

\begin{abstract}
  Large Language Models (LLMs) are increasingly used to generate and edit scientific abstracts, yet their integration into academic writing raises questions about trust, quality, and disclosure. Despite growing adoption, little is known about how readers perceive LLM-generated summaries and how these perceptions influence evaluations of scientific work. This paper presents a mixed-methods survey experiment investigating whether readers \reedit{with ML expertise} can distinguish between human- and LLM-generated abstracts, how actual and perceived LLM involvement affects judgments of quality and trustworthiness, and what orientations readers adopt toward AI-assisted writing. Our findings show that participants struggle to reliably 
  identify LLM-generated content, yet their beliefs about LLM involvement significantly shape their evaluations. Notably, abstracts edited by LLMs are rated more favorably than those written solely by humans or LLMs. We also identify three distinct reader orientations toward LLM-assisted writing, offering insights into evolving norms and informing policy around disclosure and acceptable use in scientific communication.
\end{abstract}



\maketitle

\section{Introduction}

In early 2025, a PhD student at the University of Minnesota was expelled after being accused of using a Large Language Model (LLM) to generate answers for an exam \citep{Gerezgiher2025UMNexpulsion}. 
The university’s determination rested on faculty reviewers’ judgments that parts of the student’s exam answers showed stylistic and structural similarities to answers produced by ChatGPT. The student, however, maintained that he had only used AI tools for grammar checking and denied generating any exam content with ChatGPT.
The case drew national headlines not only because of the severity of the punishment, which is striking given that detecting LLM-generated text is far from certain, but also because it reveals a broader tensions surrounding AI use in academia and scientific communities.
As LLMs become increasingly capable at complex tasks like summarization \citep{zhang2023benchmarkinglargelanguagemodels,askari-etal-2025-assessing}, paraphrasing \citep{wahle-etal-2022-large}, and translation \citep{peng-etal-2023-towards}, more scholars are adopting them into their research processes \citep{liang2024mappingincreasingusellms,Prillaman2024ChatGPTHyperProductive}. A 2024 survey of 816 academic authors across disciplines and found that 81\% had already used LLM tools in some aspect of their workflow \citep{liao2024llmsresearchtoolslarge}. Analyses of recent OpenReview data suggest that between 6.5\% and 16.9\% of submitted peer reviews may already use LLMs \citep{zhou-etal-2024-llm,liang2024monitoringaimodifiedcontentscale}.
Yet, there is no consensus on what constitutes appropriate use and how, if at all, LLM use needs to be disclosed. Policies are mixed, with many publication venues prohibiting blind text generation (e.g., AAAI, AIES), some explicitly permitting LLM use for editing or polishing (e.g., CHI, CSCW, Neurips), and others not addressing the issue at all.
The lack of consensus in policies is particularly salient when it comes to research summaries, which are both widely read and increasingly subject to LLM involvement.

Research summaries, and especially abstracts, are central to how scholarship is evaluated, shared, and cited. They function as the primary entry point for readers deciding whether to engage with a paper and are expected to be trustworthy, concise, accurate, and comprehensive. As LLMs become increasingly capable of generating or editing abstracts, concerns about how such texts are perceived and trusted become increasingly important.

At the same time, research summaries written with LLM assistance hold considerable promise. If they can be made trustworthy, they could broaden access to science by producing lay summaries for students, policymakers, and the public. They may also support interdisciplinary understanding, reduce linguistic barriers through multilingual summarization, and reduce the workload for authors drafting abstracts. Realizing this potential requires a careful understanding of how readers perceive LLM-generated summaries and how those perceptions shape judgments of quality and trust.


Researchers have documented factual errors, stylistic vagueness, and overgeneralization in LLM-generated summaries \citep{Tang2023EvaluatingLL,Catherine_gpt_summary,peters2025generalizationbiaslargelanguage}.
Yet we still lack systematic evidence on how readers themselves perceive and trust such summaries in academic contexts. \citet{HADAN2024100095} made some leeway on this question by studying reviewers’ ability to identify and evaluate AI-generated writing in the peer-review process, but this remains a single recent study focused specifically on reviewers, highlighting the need for additional investigation into general reader perceptions.
More research is required to understand whether readers can reliably distinguish LLM- from human-written abstracts, and how actual versus perceived involvement of LLMs shapes judgments of quality and trust. Moreover, little is known about the orientations that readers adopt toward LLM-assisted writing and how these stances manifest in evaluation practices. Addressing these gaps is crucial for guiding norms of disclosure, informing policy on acceptable use, and realizing the benefits of LLM-assisted summarization without undermining trust in scientific communication.

To address these gaps, we pose the following research questions:
\begin{itemize}
    \item [\textbf{RQ1}] Can participants accurately distinguish between LLM-generated and human-generated research summaries?
    \item [\textbf{RQ2}] How does the actual degree of LLM involvement (i.e., fully LLM generated, human-written, or human-written and LLM edited) affect participants’ perceptions of information quality and trustworthiness?
    \item [\textbf{RQ3}] How do participants’ beliefs about LLM involvement, regardless of their accuracy, influence their assessments of quality and trust?
    \item [\textbf{RQ4}] What orientations toward LLM-generated content emerge among participants, and how are these manifested in their evaluation practices?
\end{itemize}
\redelete{We address these questions through a mixed-design survey experiment that combines quantitative ratings with qualitative analysis of participants’ reasoning.
Our study shows that participants cannot reliably distinguish between LLM- and human-written abstracts, but their perceptions of trust and quality are strongly shaped by their beliefs about LLM involvement. We also find that abstracts that were written by human authors, either in collaboration with an LLM or without, are evaluated more favorably than fully LLM-generated ones, and we identify three distinct orientations that participants adopt toward AI-assisted writing.} \reedit{We address these questions through a mixed-design survey experiment combining quantitative ratings, behavioral measures, and LLM-in-the-loop qualitative analysis.
Because our corpus draws on ML papers readily available through arXiv, we recruited readers with corresponding ML expertise to assess them. Thus, our findings capture perceptions within this community.
This paper makes four key contributions:

\begin{itemize}
    \item [(1)] An empirical evaluation of readers’ ability to detect LLM involvement in research abstracts, showing that participants cannot reliably distinguish human- from LLM-generated or LLM-edited summaries, even though they express high confidence in their assessments.
    \item [(2)] A comparison of how actual LLM involvement shapes perceptions of quality and trust, demonstrating that LLM-edited (human-written and LLM-revised) abstracts are evaluated most favorably across clarity, conciseness, and trustworthiness, while fully LLM-generated abstracts receive the lowest evaluations.
    \item [(3)] Evidence that beliefs about LLM involvement meaningfully influence evaluation. Authorship perceptions affect judgments of clarity, comprehensiveness, engagement, and credibility, revealing a human-authorship preference and distinct evaluation heuristics under uncertainty.
    \item [(4)] Identification of three reader orientations toward AI-assisted writing: Disclosure Advocates, Pragmatic Skeptics, and Optimists. These orientations capture attitudes toward LLM-generated content, explain variation in evaluation practices, and showcase emerging norms around disclosure, acceptability, and trust.
\end{itemize}
}

\section{Related Work}

\subsection{Trust, Information Quality, and Measurement}

Trust is often described as ``the attitude that an agent will help achieve an individual’s goals in a situation characterized by uncertainty and vulnerability'' \cite{Lee2004}.
In the context of AI, trust captures users’ attitudes toward relying on model outputs under uncertainty, distinct from reliance (behavior) and trustworthiness (system quality).

\redelete{AI-assisted writing, studies} \reedit{Studies of AI-assisted writing} find that language models improve surface quality such as grammar and fluency but may reduce depth or originality \citep{HADAN2024100095,chatgpt_essays}. These mixed effects make trust a key outcome for understanding how readers evaluate AI-authored or AI-edited text.

Trust has been measured through behavioral and self-report metrics, including reliance calibration and weight-of-advice approaches \cite{10.1145/3544548.3581058,10.1145/3519266,10.1145/3544548.3581197,10.1145/3544548.3581393}. Across studies, performance and perceived quality are among the strongest predictors of trust \citep{9-yin-2019}. Transparency and explanations can strengthen trust but show mixed effects on calibrated reliance \citep{12-valeriechen-2023}. Individual differences also matter: users with more prior exposure to AI often exhibit higher trust, whereas those with deeper technical expertise may be more cautious \citep{user_attributes_trust}. \reedit{To answer RQ2 and RQ3, we evaluate abstracts along established dimensions such as clarity and comprehensiveness, aligning our measures with prior work on trust and information quality \cite{13-rae-2024, 8-Liu2023ConsumerHI, 10.1108/EJM-02-2020-0084}.}

\subsection{Trust in LLM-Generated Text and Summaries}

Recent work highlights concerns for trust in LLM-generated summaries. \citet{huang2024trustllmtrustworthinesslargelanguage} establishes a benchmark for trustworthiness, emphasizing that it is multi-dimensional, involving not only accuracy but also robustness and fairness. Several empirical studies have evaluated LLMs directly in summarization tasks \cite{zhang2025comprehensivesurveyprocessorientedautomatic, 10.1145/3731445}. 
In medical evidence summarization, \citet{Tang2023EvaluatingLL} find that models can produce factually inconsistent summaries, sometimes overstating results and other times introducing ambiguity that misrepresented the evidence. 
\citet{Catherine_gpt_summary} test ChatGPT on generating medical abstracts from only titles and journal-style cues. They show that while AI detectors could identify most outputs, human reviewers misclassified a substantial fraction and described them as vaguer and more formulaic. 
In academic writing, \citet{peters2025generalizationbiaslargelanguage} compare LLM summaries of full abstracts and articles to original texts and expert digests. They find the LLMs' tendency to exaggerate claims and overgeneralize. Although this study did not measure trust directly, such amplification poses a risk that readers could accept overstated results as faithful representations. 
Beyond factual quality, trust is also shaped by perceptions: \citet{HADAN2024100095} find that peer reviewers often judged AI-augmented submissions as more polished but less substantive, misidentified whether AI had been used, and reduced credibility when AI authorship was suspected. Together, these findings show that inconsistency, vagueness, overgeneralization, and skeptical human perceptions all shape how readers and reviewers may misinterpret or mistrust LLM-generated research writing.
\reedit{To build on these findings, we distinguish among three common forms of LLM involvement in scientific abstracts (i.e. human-written, LLM-generated, and LLM-edited) to examine whether they elicit different perceptions of information quality and trust (RQ2).}


\subsection{Human and Algorithmic Detection of LLM Involvement}  
Studies show that people often struggle to distinguish AI from human content. 
In art, \citet{10.1145/3334480.3382892} find modest ability to recognize authorship, while \citet{köbis2020artificialintelligenceversusmaya} report that detection depended on whether AI outputs were pre-selected or random. In text, accuracy often hovers near chance. \citet{Jakesch2023} and \citet{clark-etal-2021-thats} both find participants can not reliably distinguish human and LLM outputs. Instead, intuitive but flawed heuristics are used for detection (e.g. assuming the use of personal pronouns signals a human author) \cite{Jakesch2023}. 
\citet{HADAN2024100095} survey HCI peer reviewers and found that they
frequently misclassify the authorship of abstracts with accuracy that did not exceed random chance significantly. As models improved, human detection became even harder \cite{13-rae-2024}.
Researchers developed detectors to address this challenge. DetectGPT \cite{detectgpt} introduced a zero-shot perturbation-based approach, while Ghostbuster \cite{verma-etal-2024-ghostbuster} used ensemble features for cross-domain generalization. GPT-who \cite{venkatraman-etal-2024-gpt} leverages psycholinguistic principles and information density, outperforming earlier systems. Yet accuracy remains limited: \citet{FIEDLER2025100321} find that even experts and state-of-the-art detectors perform only slightly above chance, with professional-level AI text especially difficult to classify. Reliable authorship attribution thus remains elusive.
\reedit{To examine whether these attribution challenges extend to scientific abstracts, we include an authorship-guess task (RQ1) and relate these judgments to readers' evaluations.}

\subsection{Effects of Disclosing AI Involvement}
Across domains, disclosure of AI involvement often lowers trust and appreciation. \citet{parshakov2025usersfavorllmgeneratedcontent} find that participants prefer AI-written answers when the source was hidden, but this preference drops once AI authorship was revealed. Similarly, \citet{Zhang_Gosline_2023} show that AI-generated ads rate higher in quality than human ones, yet disclosure reduced this advantage, reflecting pro-human bias. \citet{zhu-etal-2025-human} documented strong label effects: identical texts labeled ``Human'' were rated far more positively than those labeled ``AI''.
Disclosure also undermines perceptions of creators. Across 13 experiments, \citet{SCHILKE2025104405} found that AI disclosure consistently reduces trust in individuals and organizations, driven by lower perceived legitimacy. \citet{13-rae-2024} confirm this in written content: although trust in text quality did not vary, creators labeled as using AI were seen as less qualified and effortful.

Domain-specific findings echo this trend. \citet{1-walker-2024} study student trust in ChatGPT versus experts. Students preferred correct AI answers when sources were hidden, but labeled conditions revealed pro-human bias. Individual attitudes moderate these effects: people with positive views of AI or confidence in its accuracy showed smaller trust declines \cite{SCHILKE2025104405}, while demographics played little role. \reedit{This variability suggests that disclosure effects may depend on broader orientations toward AI use, motivating our examination of reader profiles in RQ4.}
\citet{Hosseini2023UsingAT} raises concerns about using AI to produce scholarly publications, warning that credibility and authorship norms may be undermined without transparent disclosure.
Overall, while transparency is ethically important, disclosure often triggered skepticism, lowering appreciation of content and trust in creators. \redelete{Managing disclosure carefully is still being discussed.}
\reedit{To test how these dynamics apply to scientific abstracts, we manipulate whether LLM involvement is disclosed, enabling us to separate effects of actual versus perceived authorship (RQ3).}
\section{Methods}

\subsection{Study Design}
We designed an online survey-based experiment in which participants complete a series of three tasks evaluating three abstracts per task. Within a task, abstracts correspond to the same research paper but differ in authorship with the following options: the original human-written version (\emph{human-written abstract}), a version generated from the full research paper by an LLM (\emph{LLM-generated abstract}), and a version edited by an LLM based on the original human-written abstract (\emph{LLM-edited abstract}). Each task uses one of the abstract types as the focal target of evaluation while the other two versions are used for comparison. The order of abstract types is counterbalanced using a Latin square design \reedit{\citep{Grant1948}} to mitigate ordering effects and provide within-subject variation.

To assess how awareness of LLM involvement influences peoples' judgments, we add a between-subjects component to the experiment which ultimately contributes to a mixed study design. Participants are randomly split into two conditions: An \emph{information condition}, where the authorship of each abstract is disclosed, and a \emph{guess condition}, where participants are asked to gauge the degree of LLM involvement without additional information. 

Across all tasks and conditions, participants are asked to rate abstracts across various Likert scales related to information quality and trust. Participants are prompted to explain their choices in several free-text fields and, when presented with all three abstract types at once, are asked to identify their preferred abstract for each paper. Along with the main task-specific information, we also ask participants to provide feedback via post-task and post-survey questionnaires.

\subsection{Participants}

\begin{table}[t]
\centering
\caption{\reedit{Participant demographics (N = 69)}.}
\label{tab:participants}
\begin{tabular}{@{}ll@{}}
\toprule
\textbf{Category} & \textbf{Value} \\ 
\midrule
Gender & 23.2\% Female; 73.9\% Male; 2.9\% Prefer not to say \\[3pt]
Age groups & 25--34: 37.68\%; 34--55: 52.17\%; Other: 10.15\% \\[3pt]
English proficiency & 44.93\% native; 55.07\% fluent non-native \\[3pt]
Education level & 81.16\% doctoral or equivalent \\[3pt]
Years since degree & >5 years: 43.48\%; $\leq$ 5 years: 49.28\%; In progress: 7.25\%\\[3pt]
Professional background & Industry: 59.42\%; Academia: 30.43\%; Other: 10.15\% \\[3pt]
LLM familiarity & Little/none: 2.9\%; Some: 26.09\%; Good: 37.68\%; In-depth: 33.33\% \\[3pt]
LLM usage & 100\% have personally used LLMs \\
\bottomrule
\end{tabular}
\end{table}

We recruited participants from two sources: (1) external participants via a professional annotation vendor, who met two criteria: completion of at least one college-level statistics or computer science course, and having submitted or reviewed an academic paper for a machine learning conference or journal in the last five years, and (2) internal participants from the authors’ institution, who met only the first criterion given their daily exposure to ML as employees of a large technology company. 
All participants received compensation. 
\reedit{The final sample included 69 participants (53 external, 16 internal). Most participants were aged 25–55. Over 80\% held or were pursuing a doctoral degree, and participants represented both industry (about 60\%) and academia (about 30\%). Nearly all reported good or expert-level understanding of LLMs and personal experience using them. More details on participant demographics can be found in Figure~\ref{tab:participants}.}

\redelete{Our final sample included 69 participants (53 external, 16 internal): 16 women, 51 men, and 2 who preferred not to say. Most were aged 25–55, with 45\% native English speakers and the rest fluent. Over 80\% held or were pursuing a doctoral degree, and participants represented both industry (about 60\%) and academia (about 30\%). Nearly all reported good or expert-level understanding of LLMs and personal experience using them. More details on participant demographics can be found in Figure~\ref{tab:participants}.

Overall, participants showed strong expertise in evaluating ML research and high familiarity with LLMs, making them well-suited for studying perceptions of LLM-generated content.}

\subsection{Materials}
\subsubsection{Paper Selection and Abstract Data}
Academic abstracts often adhere to a standardized, information-dense format, enabling systematic evaluation with and without LLM involvement.
In this study, we focus on machine learning research papers from arXiv (using category \texttt{cs.LG}), spanning October 2019 to October 2022.
The time window was chosen to ensure that papers reflect modern abstract-writing practices while predating the public release of ChatGPT which is intended to minimize the likelihood of LLM involvement in the original abstracts.
In order to maintain a high quality of papers, we only consider preprints that underwent peer-review.
From this pool of papers, we sample a fixed set of 150 research papers, the abstracts of which serve as the human-written abstracts in our survey. The number of papers was deliberately kept small to enable manual correction of typographical artifacts like inconsistent LaTeX math mode rendering in LLM-generated abstract versions (e.g. \verb|beta| instead of \verb|$\beta$|).

\subsubsection{LLMs, Prompts, and Abstract Types}

\reedit{
We generate alternative abstract versions using prompts administered to Llama 3.1 8B \citep{meta-llama-3-1-2024}. This model is selected for its open accessibility and strong quality and efficiency tradeoff, which supports reproducible experiments on commodity hardware. All generations use the default temperature of 0.5 and a maximum generation length of 1,838 tokens which corresponds to the average number of tokens in our set of human written abstracts plus two standard deviations.
To provide a common structure across all outputs, we ground synthetic abstracts in the ACM abstract guidelines \citep{acm-author-guidelines}:
\begin{itemize}
\item[] \texttt{The abstract should be 150 to 200 words and consist of short, direct, and complete sentences. It should be informative enough to serve in some cases as a substitute for reading the paper itself. It should state the objectives of the work, summarize the results, and give the main conclusions, but omit future plans and citations.}
\end{itemize}
For each paper, we consider three abstract types. First, the original abstract created by the authors (\emph{Human-written}). Second, an LLM generated abstract that is written entirely by the model (\emph{LLM-generated}). Third, an LLM rewritten abstract in which the model edits the original human-written abstract (\emph{LLM-rewritten}. These three types represent realistic use cases from fully automated summarization to more collaborative forms of writing.

For LLM-rewritten abstracts, the model receives the original human written abstract together with the prompt:
\begin{itemize}
\item[] \texttt{Abstract guidelines: [ACM INSTRUCTIONS] \\
Rewrite this Abstract for a scientific paper with title '[PAPER TITLE]' based on the given abstract guidelines. Assume you are the author of the paper and start with 'Abstract:'.\\
Abstract: [ORIGINAL ABSTRACT]}
\end{itemize}

For fully LLM generated abstracts, we extracted all paper sections from the PDF, removed the original abstract, and provided the remaining paper text to the model:
\begin{itemize}
\item[] \texttt{Abstract guidelines: [ACM INSTRUCTIONS] \\
Summarize this Paper into an Abstract for a scientific paper with title '[PAPER TITLE]' based on the given abstract guidelines. Assume you are the author of the paper and start with 'Abstract:'.\\
Paper: [PAPER TEXT]}
\end{itemize}

In both conditions we select all text following the marker \texttt{Abstract:} as the final abstract. As a final step, all outputs receive minimal post processing by one of the authors to correct small typesetting issues such as missing \LaTeX{} math delimiters. No substantive edits were made.

Brief text analysis confirms that our generated and rewritten texts exhibit characteristic linguistic differences while remaining broadly comparable to human-written abstracts (Appendix~\ref{sec:text_analysis}).
}

\redelete{
\subsubsection{Textual Characteristics of Abstract Types}
\label{sec:text_analysis_main}
We compare the three abstract types (human-written, LLM-generated, and LLM-rewritten) on basic textual and linguistic properties to contextualize participant judgments (full details in Appendix~\ref{sec:text_analysis}). Human-written abstracts were slightly longer and more homogeneous, reflecting conventionalized writing practices, whereas LLM-generated and especially LLM-rewritten abstracts were shorter and more lexically diverse. Across types, most common words overlapped, but abstracts with LLM involvement were more likely to include terms such as `novel' and `approach', echoing prior work associating these with LLM authorship \cite{liang2024monitoringaimodifiedcontentscale,liang2024mappingincreasingusellms}. These findings confirm that our generated and rewritten texts exhibit characteristic linguistic differences while remaining broadly comparable to human-written abstracts.
}

\subsection{Measures and Survey Procedure}
\label{measures_survey_procedures}

\newcolumntype{L}[1]{>{\raggedright\arraybackslash}p{#1}} 
\newcolumntype{C}[1]{>{\centering\arraybackslash}p{#1}}   

\begin{table}[t]
\centering
\begin{small}
\begin{tabularx}{\textwidth}{@{}L{2.5cm} C{2cm} C{2.1cm} L{7.5cm}@{}}
\toprule
\textbf{Dimension} & \textbf{Cronbach's $\alpha$} & \textbf{Factor Loading} &\textbf{Associated Survey Questions} \\
\midrule
\textbf{Trust} & 0.910 &
0.380 \newline
0.355 \newline
0.601 \newline
0.977 \newline
1.020 \newline
0.713 \newline
&
\textit{The abstract presents logical and plausible information.} \newline
\textit{The abstract is free from ideological or factual bias.} \newline
\textit{The abstract is of high quality.}\newline
\textit{The abstract is trustworthy.}\newline
\textit{Based on the abstract, I trust the research.}\newline
\textit{Based on the abstract, I expect the research to be of high quality.}
\\
\smallskip
\textbf{Clarity} &0.830 &

0.314\newline
1.036\newline
0.696\newline
0.650\newline
&
\textit{The abstract includes only essential and relevant information.} \newline
\textit{The abstract is free from formatting or typographical errors.}\newline
\textit{The abstract is structured logically and flows smoothly.} \newline
\textit{The abstract uses clear language.} \\
\smallskip
\textbf{Comprehensiveness} & 0.879 &
0.752\newline
0.391\newline
0.973\newline
0.999\newline
&
\textit{The abstract allows readers to grasp the essence of the research.} \newline
\textit{The abstract clearly states the objectives.} \newline
\textit{The abstract effectively communicates the main conclusions.} \newline
 \textit{The abstract provides a clear summary of the results.}\\
\smallskip
\textbf{Engagement} & 0.856 &
0.533\newline
0.796\newline
0.913\newline
&
\textit{Based on the abstract, I believe the research's findings are promising.}\newline
\textit{Based on the abstract, I think the research is relevant.} \newline
\textit{Based on the abstract, I am interested in reading the full paper.}
\\
\smallskip
\textbf{Conciseness} & -- &
0.977\newline
&
\textit{The abstract has an appropriate length.} \\
\bottomrule
\end{tabularx}
\end{small}
\caption{Information quality dimensions, associated survey questions (Likert scale: –2 to +2), and question factor loadings onto the dimensions according to Exploratory Factor Analysis (EFA). Cronbach’s $\alpha$ with values $>0.8$ indicates high internal consistency for each multi-item scale. In addition to core quality dimensions such as Credibility, Clarity, and Comprehensiveness, we include Engagement as a user-centered dimension reflecting interest and motivation.}
\label{tab:survey_dimensions}
\end{table}

\subsubsection{Procedure}
Participants complete an online survey consisting of three main tasks. In each task, they are exposed to a combination of paper title and abstract, and answer a series of questions designed to assess their perceptions of LLM involvement, quality, and trustworthiness. Depending on the condition assignment, participants either evaluate the abstract without knowing the extend of LLM involvement (guess condition) or they are told explicitly which of the three types of abstracts is shown to them (information condition). We then reveal all three abstract types for the same paper and solicit participants' preferences and the reasoning behind them.

We conducted two rounds of pilot testing with internal participants. The pilots helped us to understand (1) whether the instructions are easily understandable, (2) whether the flow of the survey feels natural, (3) how long the survey takes and how many tasks each participant can complete, and (4) whether individual wordings of questions are clear. 
The final survey was designed to take approximately 60 minutes, and participants were asked to refrain from using LLMs.  

Abstract types are rotated using a Latin square design \reedit{\citep{Grant1948}}, and participants are evenly split into the information groups leading to an mixed study design.
The survey data is comprised of numerical, ordinal and free-text responses that, in conjunction with the raw text data from the abstracts, is used to address our research questions.
We describe the most important evaluation measures below and refer to Appendix~\ref{app:full_survey} for the full survey instrument.

\subsubsection{LLM authorship attribution}
To assess whether readers can detect LLM involvement, half of the participants were assigned to a guess condition in which the source of each abstract (human-written, LLM-generated, or LLM-rewritten) was not disclosed. After reading each task's focal abstract, participants were asked to gauge the likelihood that an LLM had been involved in any way to write the abstract on a 5-point Likert scale from 0=Not at all likely to 4=Extremely likely. They were then asked to guess whether the abstract was human-written, LLM-generated, or LLM-rewritten. Confidence in the assessment was indicated on a similar 5-point Likert scale, and we ultimately asked to provide reasoning for the decisions via a free-text field. We use the collected data to study if the LLM authorship attributions are more accurate than guessing at random, and how the ability to determine LLM involvement changes with the true degree of LLM involvement. 

\subsubsection{Information Quality, Trust, and Composite Score Construction}



To assess participants' perceptions of abstracts, we designed 18 survey questions covering information quality dimensions, e.g. clarity, conciseness, comprehensiveness, credibility, and objectivity. These dimensions were selected based on prior literature \cite{13-rae-2024, 8-Liu2023ConsumerHI, 10.1108/EJM-02-2020-0084} as well as pilot studies identifying salient aspects of abstract evaluation by researchers. All questions were presented as positively phrased statements like `The abstract uses clear language', and participants were asked to rate each item on a 5-point Likert scale (-2 = Strongly Disagree to 2 = Strongly Agree).

We conduct exploratory factor analysis (EFA) \reedit{\citep{Tabachnick2012-nj}} using weighted least squares extraction method and promax rotation to identify latent dimensions in participants' ratings. The analysis revealed a five-factor solution accounting for 69.18\% of the total variance. Table~\ref{tab:survey_dimensions} depicts factor loading results, associated survey questions, and Cronbach's alpha values. \reedit{Here, Cronbach’s alpha reflects the internal consistency of the items that make up each factor and indicates how reliably those items measure the same underlying construct.} All identified constructs reached alpha values of $>0.8$ suggesting high internal consistency.
The resulting dimensions were Trustworthiness, Clarity, Comprehensiveness, Engagement, and Conciseness.

We use the EFA results to create composite scores for each factor dimension by averaging ratings across the associated questions. Each question is associated to the dimension with highest factor loading using a $>0.3$ cutoff. These scores capture participants' overall judgments of abstracts along meaningful dimensions while reducing item-level noise and redundancy. We use the composite scores as outcome variables to study how perceived or de-facto LLM involvement in abstract generation impacts perceptions of information quality and trustworthiness. 

\subsubsection{Behavioral Trust}

Behavioral trust matters because it shows how users actually rely on AI in practice \cite{6-schemmer2022ifollowaibasedadvice}.
After judging individual focal abstracts for a paper, participants' are shown all three versions of abstracts for the same paper: Human-written, LLM-generated, and LLM-rewritten. They are then asked to describe the main differences between the three abstracts in a free-text field, and to choose which of the abstracts they prefer. We ask to explain their choice and, if given the chance, what aspect they might want to change about the selected abstract. We use the participants' responses to these questions to further analyze how participants think and reason about abstract quality. The selected abstract options are coded by abstract type and used as outcome variable to model behavior based preferences for LLM involvement. We conceptualize these preferences as behavioral indicators of comparative trust.

\subsubsection{Orientations Toward LLM-Generated Content}
Beyond task-specific judgments, we also examine how participants’ broader orientations toward LLMs shape their evaluation practices. Participants rated a set of statements about their prior experiences with LLMs, beliefs about their capabilities and limitations, views on the need for disclosure of LLM use, and opinions about the role of LLMs in research. Exact wordings of the statements can be found in Appendix~\ref{app:full_survey}.
In order to avoid priming participants’ experiment responses with explicit attitude questions about LLMs, this set of questions was administered at the end of the survey. Responses were given on a 5-point Likert scale of agreement. We use these responses to derive participant profiles that capture distinct orientations toward LLM-generated content.

\subsection{Analysis Approach}

\subsubsection{Quantitative Analysis}
We conduct our analysis using survey data including binary, numeric Likert-scale, multiple and single-choice, as well as free-text fields.
\redelete{Whenever useful we draw on the textual abstracts to make comparisons between participants' perceptions and objective characteristics (e.g., length of abstracts or specific linguistic features).}
For any in-between and within-subject comparisons, we use linear mixed effects models controlling for participant IDs as random effect and task IDs as fixed effects. \reedit{Task IDs are included to capture potential fatigue or learning effects, i.e. changes between participants' judgments between the first, second, and third task they are completing.}
\reedit{Because the participant sample is relatively small, we cannot robustly model additional covariates like demographics or prior experience with LLMs without compromising the stability of the estimates.}
We set the significance threshold at 0.05 and apply Holm–Bonferroni correction to adjust for multiple comparisons.
We excluded a small number of participant responses (2 tasks in total) since participants recognized the paper \redelete{which could bias their responses}\reedit{, and may thus have external information on which of the abstracts is the original human-written abstract}. Our six Latin square blocks \reedit{\citep{Grant1948}}, each corresponding to a a unique abstract type order rotation and condition (guess vs. information), have 30 to 39 participants. \reedit{Since each participant completes 3 tasks, this translates to 98 tasks in the guess condition and 107 tasks in the information condition.} Repeated measures and sample size imbalances are accounted for by the use of linear mixed-effects models.

\subsubsection{LLM-in-the-Loop Thematic Analysis}

We conduct qualitative analysis of open-ended responses using an LLM-in-the-loop thematic analysis approach adapted from~\citet{dai-etal-2023-llm}. This method combines the efficiency of LLM-assisted coding with human oversight to ensure validity and interpretive depth. Our inductive analysis proceeds through multiple stages. We begin with \textbf{initial exemplar generation}: Using Claude Sonnet 3.5 V2, we generated initial codes from a subset of participant responses, identifying recurring patterns and creating "how-to" exemplars for consistent coding. This was followed by \textbf{human validation}: The same human researcher reviewed all exemplars for accuracy, relevance, and authentic representation of participant perspectives. We then moved to \textbf{full dataset coding}: The validated exemplars guided coding of the complete dataset using Claude Sonnet 3.5 V2, with iterative prompt refinement based on quality checks. To maintain rigor, we implemented \textbf{quality assurance}: we randomly sampled 10\% of coded data for human verification, refining system prompts until achieving satisfactory coding quality. The process concluded with \textbf{thematic synthesis}: Claude Opus 4.1 synthesized low-level codes into higher-order themes, with human researchers verifying the final thematic structure and selecting representative quotes. All human input to the LLM-assisted qualitative coding was given by the same researcher 

We applied this core method to two sets of questions with adaptations for their specific analytical needs. For individual abstract evaluation (Q23), where participants described peculiarities they noticed in abstracts, we directly applied the method to all responses. Next, for comparative abstract analysis (Q25-28), which captured how participants compared and selected between abstracts, we extended the method with additional preprocessing steps. First, we separated responses by experimental condition (information vs. guess), as we are looking to understand how the knowledge of authorship fundamentally altered how participants approached evaluation. Second, we used LLM-assisted extraction to identify positive and negative evaluative statements within each response. This separation revealed what participants explicitly valued versus criticized in their abstract selections. We then applied our core thematic analysis separately to these preprocessed datasets—analyzing positive and negative statements for each condition independently. This approach allowed us to systematically compare how evaluation lenses shifted between conditions. Additionally, we conducted a second analysis focused on abstract types rather than conditions. We extracted all evaluative statements about human-written, LLM-generated, and LLM-edited abstracts from both conditions and analyzed them by abstract type.
\section{Results}

\subsection{Participants' Ability to Identify LLM-Generated Summaries (RQ1)}
\label{rq1}



\begin{table}[ht]
\centering
\begin{tabular}{lccc}
\toprule
\textbf{True Label} & \shortstack{\textbf{Guessed:}\\\textbf{LLM-Generated}} & \shortstack{\textbf{Guessed:}\\\textbf{Human-Written}} & \shortstack{\textbf{Guessed:}\\\textbf{LLM-Edited Human}} \\
\midrule
LLM-Generated Abstract     & 3  & 20 & 10 \\
Human-Written Abstract     & 3  & 14 & 15 \\
LLM-Edited Human Abstract  & 4  & 11 & 18 \\
\bottomrule
\end{tabular}
\caption{Confusion matrix of participant guesses versus true abstract types.}
\label{tab:confusion_matrix}
\end{table}

We evaluate participants' ability to correctly attribute the degree of LLM involvement in paper abstracts using data from a total of 98 tasks over 33 participants from the guess condition. A mixed-effects model predicting the accuracy of participants' guesses over three abstract types, i.e. LLM generated, human-written, and human-written with LLM edits, shows that participants were not able to attribute abstract authorship better than chance (offset = log-odd of $1/3$) for any of the three abstract types.
The confusion matrix in Table~\ref{tab:confusion_matrix} reveals a general tendency of participants to believe abstracts are written with human involvement (i.e. human-written or LLM-edited human-written) rather than entirely LLM-generated which is corroborated by a mixed-effects logistic regression indicating only about 10\% of classifications as LLM-generated with no human involvement ($\hat{p}=0.10$, 95\% CI [0.06, 0.18]).

When asked directly how likely an LLM was involved in creating a given abstract, participants rated all abstract types significantly above the neutral midpoint of the scale (Human-written: $\beta=1.63$, $p<.001$; LLM-edited: $\beta=1.79$, $p<.001$; LLM-generated: $\beta=1.30$, $p<.001$). However, the differences between abstract types were not statistically significant.
Participants reported moderate confidence levels in their assessments of LLM involvement for all abstract types (Human-written: $\beta$=2.456, $p<0.001$; LLM-edited: $\beta$=2.383, $p<0.001$; LLM-generated: $\beta$=2.61, $p<0.001$). There was no evidence of task order effects in any of the models, suggesting consistency of participants' authorship attributions across the three tasks. 

Taken together, these results reveal a human preference bias in authorship attribution, with participants rarely labeling abstracts as purely LLM-generated, alongside a baseline suspicion of LLM involvement. This suggests ambiguity in attribution: readers treat human authorship as the default yet assume LLM assistance is pervasive and difficult to rule out. \redelete{To better understand the reasoning behind these attribution patterns, we analyze participants’ qualitative explanations for authorship attribution, which revealed recurring strategies targeting abstract completeness, clarity, authenticity, credibility, engagement, and writing conventions.} \reedit{To better understand the reasoning behind these attribution patterns, we qualitatively analyze participants' written justifications. Table~\ref{tab:llm_detection_strategies} presents a taxonomy of the resulting detection strategies, ranging from Completeness \& Coverage to Writing Conventions. While some heuristics aided detection, others often backfired, with participants incorrectly flagging human-written text as AI based on subjective stylistic preferences.}

\redelete{
\textbf{Completeness and Coverage \reedit{(N = 22)}:} When evaluating completeness and coverage participants primarily detected LLM-authorship by examining the presence or absence of essential research components: problem statement, methodology, findings, and implications.
They interpreted structural gaps as evidence of AI generation leading to both correct identifications and misattributions.
For example, \textit{P14} incorrectly attributed a human-written abstract to LLM involvement, stating, \textit{"There are no results given. The abstract feels like LLM generated. It does not give any detail about the proposed GIL method."} Yet \textit{P24}, using the same strategy, correctly identified LLM involvement: \textit{"The peculiar thing is the abstract mentioning how this is a novel solution without even discussing the flaws of existing research."}

\textbf{Clarity and Accessibility \reedit{(N = 16)}:} 
Participants noticed writing clarity as a signal but varied in how they used this observation.
Some participants, such as \textit{P49}, noted clarity features without explicitly connecting them to LLM-authorship. \textit{P49} found a human-written abstract \textit{"confusing and the long sentences make it hard to follow. I got lost in the process,"} while describing a generated abstract as \textit{"easy to follow... the writing style is easy and smooth."} Though \textit{P49} clearly noticed these clarity differences, they didn't explicitly use them for detection.
Other participants
directly used clarity assessments as evidence of authorship, often leading to incorrect attributions. \textit{P31} praised a generated abstract for being \textit{"very down to earth, written by a well educated researcher who has a clear understanding of the problem and can eloquently articulate it in an understandable manner,"} using perceived clarity and eloquence as evidence of human authorship. This resulted in systematic misattribution where AI-generated content was praised for distinctly human communication qualities\reedit{, echoing concerns \cite{Catherine_gpt_summary} that clear, polished writing can bias authorship judgments.}

\textbf{Authenticity and Authorship \reedit{(N = 28)}:} 
Participants showed systematic misattribution patterns, confidently assigning incorrect authorship while expressing certainty about their assessments. \textit{P12} misidentified a human-written abstract, stating \textit{"The use of unnecessarily complicated phrases (most likely generated by an LLM) is quite jarring,"} specifically referring to technical descriptions. 
\textit{P6} made a similar error, citing specific phrases like \textit{"detailed dependencies"} and \textit{"not cumulative information"} as evidence: \textit{"I assumed this was written by a non-English speaker, but because of its structure, perhaps the abstract was written by an LLM."} Conversely, participants praised LLM-generated abstracts for distinctly human qualities, with \textit{P31} describing generated content as \textit{"written by a well educated researcher who can has a clear understanding of the problem and can eloquently articulate it in an understandable manner,"} demonstrating how participants confidently attributed human authorship to AI-generated text based on perceived eloquence and expertise.

\textbf{Research Credibility \reedit{(N = 7)}:} Participants also used credibility indicators as proxy for authenticity markers, using the presence or absence of quantitative data as evidence for or against LLM-authorship. 
This heuristic led to systematic misattribution, as demonstrated by \textit{P14} who used missing results to incorrectly identify a human-written as LLM-generated: \textit{"There are no results given. The abstract feels like LLM generated. It does not give any detail about the proposed GIL method."} The same participant then criticized a generated abstract for similar issues: \textit{"there are no numbers in the abstract which makes it not trustworthy."} \textit{P2} showed the reverse pattern, correctly identifying a human-written partly because \textit{"the inclusion of numbers makes the abstract seem less likely written by a LLM." }

\textbf{Reader Engagement \reedit{(N = 11)}:} When evaluating reader engagement, participants used their level of interest and motivation as indicators of authorship. 
\textit{P19} explicitly connected lack of engagement to artificial generation when assessing a generated abstract: \textit{"To be sincere, I was off from the start of the reading as there was nothing in my opinion attractive about the abstract... It gave it away as a work written without any simulation done."} However, engagement perception varied inconsistently across abstract types. \textit{P24} found a human-written engaging, stating \textit{"The way it is constructed makes readers engaged,"} while \textit{P12} correctly identified artificial involvement in a rewritten abstract, describing it as \textit{"very abrupt and robotic — this type of writing produces no desire to read the paper whatsoever."} In another example, \textit{P30} found substantive value in generated content, praising a generated abstract for \textit{"providing a new exciting angle to the field of machine learning"} based on research novelty rather than writing quality. These mixed responses suggest that engagement levels could be influenced more by research content and personal interest than by authorship type. 

\textbf{Writing Conventions \reedit{(N = 18)}:} Participants also used academic writing conventions as a way to detect LLM-involvement in the authorship. \textit{P6} criticized a human-written's language as \textit{"not good academic writing"} and concluded it was \textit{"perhaps written by an LLM."} Interestingly, \textit{P32} criticized a generated abstract for beginning inappropriately: \textit{"the beginning of the abstract stood out to me since it begins with 'This paper proposes' which I think is not an appropriate start to an abstract,"} however such openings are not uncommon in academic writing, revealing how participants' expectations about AI writing patterns led to misguided conventional assessments.}

\reedit{{
\footnotesize 
\renewcommand{\arraystretch}{1.5} 

\begin{longtable}{p{0.22\textwidth} p{0.45\textwidth} p{0.28\textwidth}}


\toprule
\textbf{Detection Strategy} & \textbf{Example Participant Reasoning} & \textbf{Example Outcome \& True Label} \\
\midrule
\endfirsthead

\multicolumn{3}{c}%
{{\bfseries \tablename\ \thetable{} -- continued from previous page}} \\
\toprule
\textbf{Detection Strategy} & \textbf{Illustrative Participant Reasoning} & \textbf{Outcome \& True Label} \\
\midrule
\endhead


\bottomrule
\endfoot

\bottomrule
\caption{Participant reasoning strategies for detecting LLM involvement, showing correct identifications, misattributions, and observations.} 
\label{tab:llm_detection_strategies} \\
\endlastfoot


Completeness \& Coverage \newline (N = 22)
& ``There are no results given. The abstract feels like LLM generated. It does not give any detail about the proposed GIL method.'' (P14) 
& \textbf{Misattribution:} Used structural gaps as evidence of AI generation. \newline (True Label: Human-Written) \\ 
\addlinespace 
& ``The peculiar thing is the abstract mentioning how this is a novel solution without even discussing the flaws of existing research.'' (P24) 
& \textbf{Correct Identification:} Recognized incomplete coverage pattern. \newline (True Label: LLM-Generated) \\
\midrule

Clarity \& Accessibility \newline (N = 16)
& ``very down to earth, written by a well educated researcher who has a clear understanding of the problem and can eloquently articulate it...'' (P31) 
& \textbf{Misattribution:} Mistook LLM fluency for human expertise. \newline (True Label: LLM-Generated) \\
\addlinespace
& ``[The human abstract was] confusing... I got lost in the process... [The AI abstract was] easy to follow... the writing style is easy and smooth.'' (P49)
& \textbf{Observation Only:} Noticed clarity differences but did not use them for explicit detection. \newline (True Label: LLM-Generated) \\
\midrule

Authenticity \& Authorship \newline (N = 28)
& ``The use of unnecessarily complicated phrases (most likely generated by an LLM) is quite jarring.'' (P12) 
& \textbf{Misattribution:} Mistook complex human writing for AI artifacts. \newline (True Label: Human-Written) \\
\addlinespace
& ``I assumed this was written by a non-English speaker, but because of its structure, perhaps the abstract was written by an LLM.'' (P6) 
& \textbf{Misattribution:} Confused non-native writing patterns with AI generation. \newline (True Label: Human-Written) \\
\midrule

Research Credibility \newline (N = 7)
& ``the inclusion of numbers makes the abstract seem less likely written by a LLM.'' (P2) 
& \textbf{Correct Identification:} Used quantitative data presence as human marker. \newline (True Label: Human-Written) \\
\addlinespace
& ``There are no results given... there are no numbers in the abstract which makes it not trustworthy.'' (P14) 
& \textbf{Misattribution:} Linked missing quantitative data to AI authorship. \newline (True Label: LLM-Generated) \\
\midrule

Reader Engagement \newline (N = 11)
& ``very abrupt and robotic --- this type of writing produces no desire to read the paper whatsoever.'' (P12) 
& \textbf{Correct Identification:} Correctly identified lack of engagement in AI editing. \newline (True Label: LLM-Edited) \\
\addlinespace
& ``To be sincere, I was off from the start of the reading as there was nothing in my opinion attractive about the abstract...'' (P19) 
& \textbf{Correct Identification:} Detected artificial generation through lack of compelling content. \newline (True Label: LLM-Generated) \\
\midrule

Writing Conventions \newline (N = 18)
& ``the beginning of the abstract stood out to me since it begins with `This paper proposes' which I think is not an appropriate start...'' (P32) 
& \textbf{Misattribution:} Mistook standard academic convention for AI pattern. \newline (True Label: LLM-Generated) \\
\addlinespace
& ``not good academic writing... perhaps written by an LLM.'' (P6) 
& \textbf{Misattribution:} Confused non-standard human writing with AI generation. \newline (True Label: Human-Written) \\

\end{longtable}
}}

\subsection{Trust and Information Quality Across Levels of LLM Involvement (RQ2)}
\label{rq3}

\subsubsection{Self-reported Trust: Participant Ratings}

\begin{table}[ht]
\centering
\small
\begin{tabular}{l|l|ll|ll|ll}
\toprule
\textbf{Outcome} & &\multicolumn{2}{c}{\textbf{LLM-Generated}} & \multicolumn{2}{c}{\textbf{Human-Written}} & \multicolumn{2}{c}{\textbf{LLM-Edited Human}} \\
                 & Task ID& Guess & Info & Guess & Info & Guess & Info \\
\midrule
Trust & $\beta=-0.039$ 
& $\beta=0.514$ & $\beta=0.544$ *
& $\beta=0.379$ & $\beta=0.776$ *** 
& $\beta=0.283$ & $\beta=0.831$ *** \\

Clarity & $\beta=-0.132$
& $\beta=0.810$ ** & $\beta=1.039$ ***
& $\beta=0.627$ * & $\beta=0.983$ *** 
& $\beta=0.821$ ** & $\beta=1.383$ *** \\

Comprehensiveness & $\beta=-0.03$
& $\beta=0.455$ & $\beta=0.559$
& $\beta=0.492$ & $\beta=0.683$ *
& $\beta=0.483$ & $\beta=0.683$ * \\

Engagement & $\beta=-0.054$
& $\beta=0.504$ & $\beta=0.647$
& $\beta=0.551$ & $\beta=0.861$ **
& $\beta=0.470$ & $\beta=0.861$ ** \\

Conciseness & $\beta=-0.027$
& $\beta=0.783$ * & $\beta=0.803$ *
& $\beta=0.746$ * & $\beta= 1.164$ ***
& $\beta=0.538$ & $\beta=0.771$ * \\

\bottomrule
\end{tabular}
\caption{Fixed effects model coefficients ($\beta$) and significance levels: $^{***}p<.001$, $^{**}p<.01$, $^{*}p<.05$.
Each row corresponds to a separate model. Positive parameters signal participants' agreement with positively worded statements (e.g., `The abstract has appropriate length'). \reedit{Task ID refers to the order in which participants completed their 3 tasks in the study.}}
\label{tab:beta_significance_table}
\end{table}

To examine the effects of abstract type conditions on information quality and trust judgments, we fit linear mixed-effects models for each of the composite outcomes introduced in Section~\ref{measures_survey_procedures}. Our models jointly estimate the influence of abstract type and information condition while controlling for task ID and participant-level random effects. Results are reported in Table~\ref{tab:beta_significance_table}. No significant fatigue or learning effects \naedit{as modeled by the task IDs} were observed\naedit{, i.e. participants' perception did not depend on whether they completed a task at the beginning of the study or in the end.}

Across all five composite outcomes, i.e. trust, clarity, comprehensiveness, engagement and conciseness, the main effects of abstract type are consistently positive or non-significant, indicating that none of the abstract types are consistently perceived to be of low quality or negative trustworthiness.
Effect sizes differ across outcomes, with the most pronounced effects observed for clarity and conciseness and less consistently significant effects for trust, comprehensiveness and engagement.
We hypothesize that clarity and conciseness in our setting are comparatively straightforward and objective to evaluate, leading to greater participant agreement, whereas trust, comprehensiveness, and engagement require deeper, more subjective evaluations beyond linguist qualities contributing to more varied results.

When the extend of LLM involvement is disclosed, participants report trust in LLM-edited abstracts ($\beta=0.831$, $p<0.001$), human-written ($\beta=0.776$, $p<0.001$) and LLM-generated ($\beta=0.544$, $p<0.01$) abstracts. 
Because our models omit the intercept, these $\beta$ coefficients represent the estimated mean trust scores for each abstract type rather than contrasts against a baseline.
Clarity is rated highest in abstracts edited by LLMs ($\beta=1.383$, $p<0.001$) or completely generated by LLMs ($\beta=1.039$, $p<0.001$), as compared to fully human-written abstracts ($\beta=0.983$, $p<0.001$); yet differences were not significant. All abstract types were judged as concise, i.e. human-written abstracts ($\beta=1.164$, $p<0.001$), LLM-generated ($\beta=0.803$, $p<0.01$) and LLM-edited ($\beta=0.771$, $p<0.01$).

We further observe that informing participants of the level of LLM involvement in the abstracts increases agreement with the quality statements in all abstract conditions.
For human-written and LLM-edited abstracts, participants consistently rate the abstracts as trustworthy, clear, comprehensive, engaging, and concise in the info condition, but many of the same parameters are not significantly different from zero in the guess condition. For LLM-generated abstracts, revealing authorship increases ratings of trustworthiness, clarity, and conciseness, but participants remain neutral on average in their assessments of comprehensiveness and engagement, regardless of condition.
Particularly interesting patterns emerge from the results for the more subjective trust, comprehensiveness, and engagement outcomes. 
Participants expressed neutral trust when uninformed about LLM involvement, but reported positive trust when authorship was disclosed. 
This was true even for fully LLM-generated abstracts, suggesting that, in our experimental setting, trust is shaped not only by whether an abstract is LLM-generated but maybe more importantly by whether the extent of LLM involvement is disclosed.

\subsubsection{Behavioral Trust: Abstract Selection}

\begin{figure}
    \centering
    \includegraphics[width=0.5\linewidth]{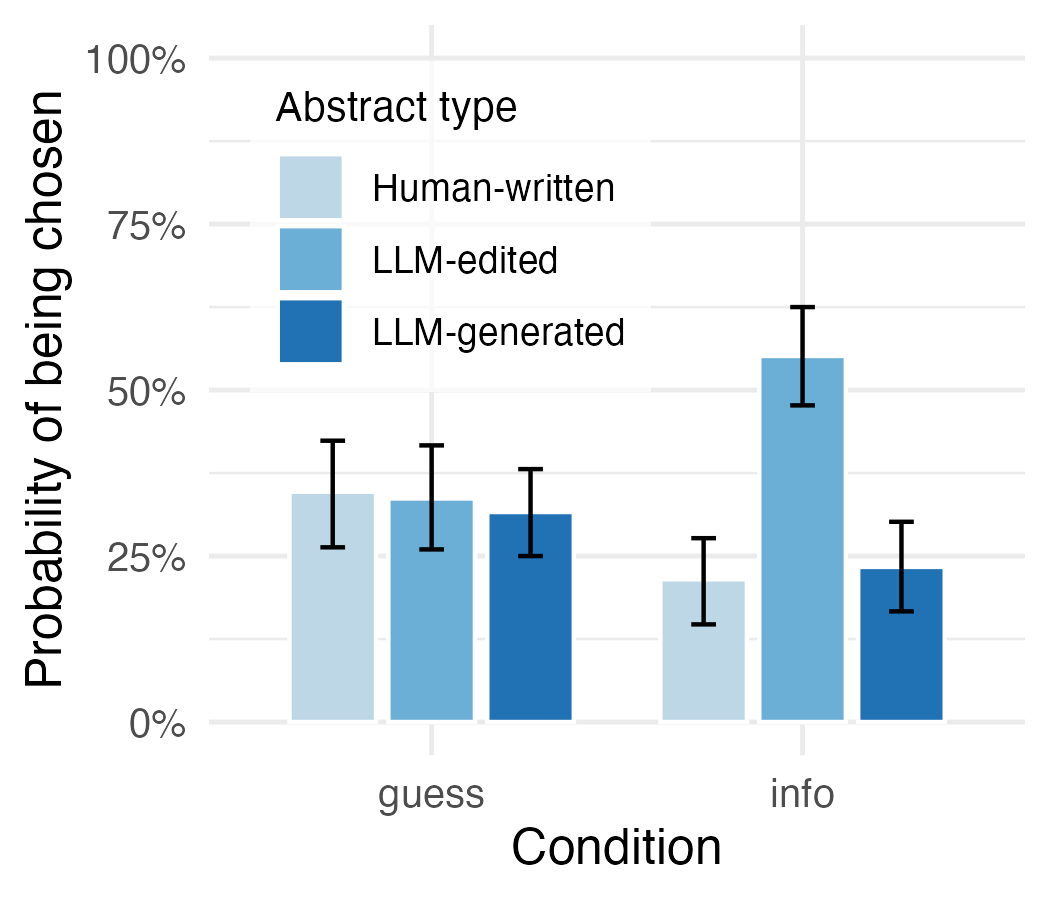}
    \caption{Probability of selecting each abstract type by information condition. Error bars indicate 95\% confidence intervals. When LLM involvement is revealed, participants show a strong preference towards LLM-edited abstracts.}
    \label{fig:behavioral_trust}
\end{figure}

In addition to self-reported ratings, we examine behavioral trust by analyzing participants’ abstract selection patterns when presented with all three types of abstracts.
We analyze results at the task level and account for the repeated measures by performing all inference using participant-clustered resampling procedures. Specifically, 95\% confidence intervals are computed using cluster bootstrapping, and hypothesis tests employ participant-level permutation tests.

Our results show that participants have an overall preference for LLM-edited abstracts (45\% vs. 27–28\% for human-written and LLM-generated, $p<0.01$).
As displayed in Figure~\ref{fig:behavioral_trust}, this preference is driven by the info condition, where over half of participants (55\%) selected the LLM-edited abstract. In fact, no significant differences in selection rates of abstract types are observed in the guess condition, consistent with the earlier observation that participants could not reliably distinguish among the abstract types without authorship information.
When controlling for the type of the first-exposed abstract, i.e. the focal abstract participants' critically engage with before being presented all three options, we find a tendency to stray from the originally shown version. In both the info and guess conditions, the first-shown abstract was less likely to be chosen than expected by chance (22\% vs. 33\%, $p < 0.001$), suggesting an over-scrutiny penalty.

\subsubsection{Understanding Abstract Preferences: A Qualitative Analysis}
To complement our quantitative results, we analyze participants’ written comparisons of the three abstract types. This reveals how readers interpret technical language, what they notice and value, and how evaluations shift depending on whether authorship was known or guessed. \redelete{Below we present themes by abstract type, highlighting differences between information and guess conditions. Section~\ref{rq4} offers a broader analysis of how authorship knowledge shapes evaluation; here we focus on qualities distinguishing each abstract type.} \reedit{Table~\ref{tab:qual_themes_abstract_pref} summarizes the key qualitative themes distinguishing each abstract type, illustrating how participants balanced factors such as structural coherence, technical depth, and authenticity signals. While Section~\ref{rq4} offers a broader analysis of how authorship knowledge shapes evaluation mechanisms, this section focuses on the specific textual qualities that drove participant preferences.}


\redelete{
\noindent\textbf{{Human-Written Abstracts.}} Four key themes emerged in how participants evaluated human-written abstracts, revealing both strengths and systematic weaknesses:

\textbf{1. Insufficient Quantitative Details \reedit{(N = 19)}:} A consistent criticism across conditions was the lack of empirical support. P40 observed \textit{``the text does not provide any quantitative elements to support the conclusions of the authors,''} while P24 noted abstracts \textit{``lack numerical values while discussing the results''} and \textit{``do not quantify the advantages over existing models.''} Multiple participants suggested adding quantitative statements to strengthen claims, highlighting this as an important weakness in human-written abstracts.

\textbf{2. Structural Coherence \reedit{(N = 8)}:} Participants consistently praised the logical organization of human-written abstracts. P42 appreciated how they \textit{``first explain the motivation, problem and then gradually flow to the solution and results,''} while P7 noted how \textit{``consistent, logically connected and direct sentences...makes the story sound more logical and trustworthy.''} P23 similarly praised \textit{``the best story-like structure: the author produces the problem, then proposes the overall solution, then highlights the contribution.''}

\textbf{3. Divisive Technical Depth \reedit{(N = 15)}:} Technical language in human-written abstracts polarized participants. Some criticized them as \textit{``Too technical - abstract jumps right into the kernel functions without describing the problem, scope etc. Too technical to no one benefit''} (P31), finding the jargon excessive. Others valued the same technical depth, with P38 praising abstracts that \textit{``achieve a strong balance between clarity and technical depth.''} This division persisted across both knowledge conditions.

\textbf{4. Writing Imperfections \reedit{(N = 15)}:} Human-written abstracts contained grammatical and stylistic flaws that participants interpreted differently based on knowledge condition. When guessing authorship, P18 saw awkward phrasing as evidence the abstract \textit{"seems like it was written by a non-english speaker or an AI that cannot structure sentences correctly."} When authorship was known, P28 simply noted \textit{"It is very short and composed of some punctuation error,"} treating these as quality issues rather than authenticity signals.
\newline

\noindent\textbf{{LLM-Edited Abstracts.}} LLM-edited abstracts were most preferred by participants overall. Four themes emerged that explain this preference and reveal both the strengths and limitations of LLM-assisted editing:

\textbf{1. Enhanced Clarity Without Sacrificing Substance \reedit{(N = 17)}:} The primary strength driving preference was improved readability that still felt authentic (human-written). When authorship was unknown, participants praised these qualities without recognizing LLM involvement, with P1 calling the writing \textit{"cohesive"} and \textit{"well-organized and easy to follow,"} and P26 noting it \textit{"strikes a balance between clarity and technical depth, making it more accessible."} When authorship was disclosed, participants explicitly attributed these improvements to LLM editing, with P15 noting \textit{"LLM introduced linguistic clarity and cohesiveness"} and P41 observing it \textit{"provides an organic flow of the ideas of the research with better choice of words."}

\textbf{2. Improved Structure and Organization \reedit{(N = 5)}:} LLM editing created clearer academic formatting that participants valued across conditions. When authorship was unknown, participants still appreciated the organization without recognizing LLM involvement. When disclosed, P27 explicitly praised these abstracts as \textit{"more structured and less complex to the reader"} while remaining \textit{"highly technical yet more structured."} These structural enhancements made complex research more digestible without oversimplification, contributing to the overall preference.

\textbf{3. Editorial Enhancement Has Limits \reedit{(N = 28)}:} LLM editing enhanced presentation but could not fix fundamental content deficiencies. P8 observed abstracts still \textit{"should include quantified results"} as \textit{"it is not clear how the performance improvements are measured and by how much is the improvement."} P45 similarly identified that abstracts \textit{"lack numerical comparison with existing study which would help grabbing attention and show clear advantages."} This limitation appeared across both knowledge conditions, highlighting a critical insight: LLM editing cannot compensate for missing empirical content in the original abstract. Authors cannot simply rely on an LLM to polish their way out of substantive gaps.

 \textbf{4. Effective Human-LLM Balance \reedit{(N = 5)}:} LLM-edited abstracts achieved the highest preference by improving clarity while maintaining substance. When participants were asked to select their preferred abstract from all three types, those who chose LLM-edited versions praised them as \textit{"well-organized and easy to follow"} (P1) and appreciated how they \textit{"strike a balance between clarity and technical depth, making it more accessible"} (P26). Even when LLM involvement was disclosed, participants valued the improvements, with P15 noting the abstract \textit{"clearly states the objective"} and presents work \textit{"in such a manner that is easy to understand to audience,"} and P41 appreciating the \textit{"organic flow of the ideas."}
\newline

\noindent\textbf{{LLM-Generated Abstracts.}} Despite strong structural qualities, fully LLM-generated abstracts received the lowest selection rates. Four themes emerged that explain both their strengths and why participants ultimately preferred other options:

\textbf{1. Information Overload Without Focus \reedit{(N = 5)}:} LLM-generated abstracts provided extensive detail but often missed essential elements. P31 criticized abstracts that \textit{"do not describe the problem,"} making it \textit{"hard to put a context to the findings,"} while P25 acknowledged they \textit{"contain more information."} When authorship was disclosed, participants identified the root issue: P4 noted \textit{"LLM-written text brings too many specific terms from the paper,"} and P10 observed it \textit{"provides a lot of technical details which are not relevant for an abstract."} This pattern suggests LLMs struggle to distinguish essential from supplementary information.

\textbf{2. Clear Structure and Organization \reedit{(N = 7)}:} The primary strength of LLM-generated abstracts was their adherence to academic conventions (which was an instruction in the system prompt). Participants consistently praised structural clarity, with P2 noting abstracts \textit{"follow standard abstract format"} and P9 confirming \textit{"the abstract is well structured and easy to follow."} P45 appreciated how \textit{"the flow is very easy to follow what is the problem statement/challenge, why this is important."} This organizational strength appeared across both knowledge conditions, suggesting LLMs effectively replicate academic formatting.

\textbf{3. Misplaced Authenticity Signals \reedit{(N = 13)}:} LLM-generated abstracts inadvertently appeared authentic through their use of conventional academic language. P18 praised generated text for feeling \textit{"like something I would easily present or say in real life"} and noted how pronouns like \textit{"We"} and \textit{"Our"} implied \textit{"ownership."} P34 similarly recognized \textit{"a more traditional academic tone with complete sentences and a clear, narrative structure."} Ironically, LLMs' successful mimicry of academic conventions made them seem genuinely human to some readers, even though these were fully machine-generated abstracts.

\textbf{4. Technical Vocabulary Without Substance \reedit{(N = 22)}:} LLM-generated abstracts included heavy technical language that obscured rather than clarified the research. P24 complained about \textit{"lots of technical jargons"} making it \textit{"complicated and tough to understand the main goal,"} while P33 noted the lack of \textit{"specifics around the results"} and wanted \textit{"summarized metrics."} This suggests LLMs mimic the surface complexity of academic writing without providing the concrete details that justify such technical presentation.
}
\reedit{{ 
\footnotesize 
\renewcommand{\arraystretch}{1.5} 

\begin{longtable}{p{0.25\textwidth} p{0.70\textwidth}}


\toprule
\textbf{Theme \& Frequency} & \textbf{Key Insight \& Illustrative Quotes} \\
\midrule
\endfirsthead

\multicolumn{2}{c}{{\bfseries \tablename\ \thetable{} -- continued from previous page}} \\
\toprule
\textbf{Theme \& Frequency} & \textbf{Key Insight \& Illustrative Quotes} \\
\midrule
\endhead


\bottomrule
\endfoot

\bottomrule
\caption{Qualitative themes explaining participant preferences for different abstract types. Themes describe what participants noticed and valued, with N indicating the frequency of the theme.} 
\label{tab:qual_themes_abstract_pref} \\
\endlastfoot


\multicolumn{2}{l}{\textbf{\textit{A. Human-Written Abstracts}}} \\ 
\midrule

\textbf{1. Insufficient Quantitative Details} \newline (N = 19) & 
Consistent criticism regarding the absence of supporting data. \newline
\textit{``The text does not provide any quantitative elements to support the conclusions...''} (P40); \textit{``[Abstracts] lack numerical values while discussing the results.''} (P24) \\
\addlinespace

\textbf{2. Structural Coherence} \newline (N = 8) & 
Praised for logical progression and narrative flow. \newline
\textit{``First explain the motivation, problem and then gradually flow to the solution and results...''} (P42); \textit{``The best story-like structure...''} (P23) \\
\addlinespace

\textbf{3. Divisive Technical Depth} \newline (N = 15) & 
Polarized views on the necessity of high technical density. \newline
\textit{``Too technical... abstract jumps right into the kernel functions...''} (P31) vs. \textit{``Achieve a strong balance between clarity and technical depth.''} (P38) \\
\addlinespace

\textbf{4. Writing Imperfections} \newline (N = 15) & 
Grammatical flaws interpreted variously as authenticity signals or quality issues. \newline
\textit{``Seems like it was written by a non-english speaker or an AI...''} (P18); \textit{``It is very short and composed of some punctuation error.''} (P28) \\

\midrule
\multicolumn{2}{l}{\textbf{\textit{B. LLM-Edited Abstracts (Most Preferred)}}} \\ 
\midrule

\textbf{1. Enhanced Clarity} \newline (N = 17) & 
Improved readability and flow without sacrificing original meaning. \newline
\textit{``Strikes a balance between clarity and technical depth, making it more accessible.''} (P26); \textit{``LLM introduced linguistic clarity and cohesiveness.''} (P15) \\
\addlinespace

\textbf{2. Improved Organization} \newline (N = 5) & 
Clearer academic formatting makes complex research more digestible. \newline
\textit{``More structured and less complex to the reader... highly technical yet more structured.''} (P27) \\
\addlinespace

\textbf{3. Editorial Limits} \newline (N = 28) & 
Polished presentation cannot mask fundamental gaps in empirical content. \newline
\textit{``It is not clear how the performance improvements are measured...''} (P8); \textit{``Lack numerical comparison... which would help grabbing attention.''} (P45) \\
\addlinespace

\textbf{4. Human-LLM Balance} \newline (N = 5) & 
Highest preference derived from combining human substance with AI clarity. \newline
\textit{``Organic flow of the ideas of the research with better choice of words.''} (P41) \\

\midrule
\multicolumn{2}{l}{\textbf{\textit{C. LLM-Generated Abstracts}}} \\ 
\midrule

\textbf{1. Information Overload} \newline (N = 5) & 
Excessive detail fails to distinguish essential from supplementary info. \newline
\textit{``LLM-written text brings too many specific terms...''} (P4); \textit{``Hard to put a context to the findings.''} (P31) \\
\addlinespace

\textbf{2. Clear Structure} \newline (N = 7) & 
Effective replication of standard academic formatting conventions. \newline
\textit{``The flow is very easy to follow what is the problem statement...''} (P45); \textit{``Follow standard abstract format.''} (P2) \\
\addlinespace

\textbf{3. Misplaced Authenticity} \newline (N = 13) & 
Mimicry of academic tone and pronouns creates false sense of human authorship. \newline
\textit{``Like something I would easily present or say in real life...''} (P18); \textit{``More traditional academic tone with complete sentences.''} (P34) \\
\addlinespace

\textbf{4. Empty Technicality} \newline (N = 22) & 
Technical vocabulary used to mask a lack of concrete substance. \newline
\textit{``Lots of technical jargons [make it] complicated...''} (P24); \textit{``Specifics around the results [are missing].''} (P33) \\

\end{longtable}
}}

\subsection{The Role of Believed LLM Involvement in Shaping Assessments (RQ3)}
\label{rq4}

\begin{table}[ht]
\centering
\small
\begin{tabular}{l|l|l|l}
\toprule
&\textbf{Believed to be} & \textbf{Believed to be} & \textbf{Believed to be}\\
\textbf{Outcome} & \textbf{LLM-Generated} & \textbf{Human-Written} & \textbf{LLM-Edited Human} \\
\midrule
Trust
& $\beta=0.443$  
& $\beta=0.316$ 
& $\beta=0.190$ \\

Clarity
& $\beta=0.590$ ** 
& $\beta=0.532$ ** 
& $\beta=0.345$  \\

Comprehensiveness
& $\beta=0.481$ * 
& $\beta=0.545$ * 
& $\beta=0.222$ \\

Engagement
& $\beta=0.434$ 
& $\beta=0.513$ * 
& $\beta=0.250$ \\

Conciseness
& $\beta=0.559$  
& $\beta=0.778$ ** 
& $\beta=0.553$  \\
\bottomrule
\end{tabular}
\caption{Fixed effects model coefficients ($\beta$) and Holm-corrected significance levels: $^{***}p<.001$, $^{**}p<.01$, $^{*}p<.05$, $^{.}p<.1$. Each row corresponds to a separate model. Positive parameters indicate greater agreement with positively worded statements.}
\label{tab:beta_significance_table_new}
\end{table}

\subsubsection{Quantitative Effects of Perceived Authorship}

In the last section, we analyzed how the degree of LLM involvement in abstracts impacts participants' quality and trust judgments. We looked at both participants who received information on LLM authorship, and participants who were left to guess LLM involvement themselves. In this section, we turn towards these guesses and examine how the the belief, rather than the reality, of LLM involvement shapes assessments. Since this analysis relies on half of the data (the guess condition), and thus far no ordering effects have been observed, we omit modeling task ID effects for this analysis. Results of our linear mixed effects models for each composite assessment score are summarized in Table~\ref{tab:beta_significance_table_new}.

Similar to the previous analyses, none of the guessed abstract types are perceived to be of systematically low quality or negative trustworthiness, as indicated by the absence of negative coefficients. Participants show a clear preference for abstracts they believe to be human-written, which are rated as especially clear ($\beta=0.532$, $p<0.01$), comprehensive ($\beta=0.545$, $p<0.05$), engaging ($\beta=0.513$, $p<0.05$), and concise ($\beta=0.778$, $p<0.01$). Believing an abstract to be LLM-generated is also associated with positive perceptions of clarity ($\beta=0.590$, $p<0.01$) and comprehensiveness ($\beta=0.481$, $p<0.05$), which may reflect common associations of AI writing with being structured, detailed, and free from ambiguity. By contrast, abstracts thought to be LLM-edited show no significant differences across outcomes, pointing to a lack of consistent perceptions of this hybrid form.
Overall, the findings point to a human preference bias: although participants could not reliably distinguish abstracts by LLM involvement, they linked qualities like engagement and conciseness to human authorship. This may reflect assumptions that humans bring greater creativity and intentionality, as well as the stereotype that LLMs tend to be unnecessarily verbose. At the same time, participants still attributed clarity and comprehensiveness when they believed an abstract to be LLM-generated, suggesting that AI authorship is associated with structure and thoroughness. Importantly, this pattern could reflect not only a bias in evaluation but also the reverse process, where participants inferred authorship from the qualities they perceived.

In contrast to the other composite scores, trust did not significantly vary by perceived authorship. On average, participants neither agree nor disagree with statements of trust across all believed abstract types. This is in line with the findings from Section~\ref{rq3} which suggest that transparency itself, rather than beliefs about or actual LLM involvement, is the key driver of trust. 

\subsubsection{How Uncertainty Transforms Evaluation: A Qualitative Analysis}
We examine participants’ explanations of how they evaluated human-written, LLM-edited, and LLM-generated abstracts under two conditions: guess (authorship unknown) and information (authorship disclosed). \reedit{As detailed in Table~\ref{tab:qual_strategies_comparison}, this comparison reveals that uncertainty led participants to use fundamentally different evaluation strategies.} Without authorship knowledge, participants operated as \textit{authenticity detectives}, focusing on surface-level stylistic cues to verify the source (Part A). However, disclosure shifted participants into the role of \textit{quality assessors}, prioritizing empirical substance and communicative effectiveness (Part B). This transformation, summarized in Part C of the table, suggests that transparency is essential for fair assessment, as uncertainty prompts readers to focus on detection rather than merit.


\redelete{
\noindent\textbf{Without Authorship Knowledge (guess condition)}
When participants had to infer authorship, their evaluations prioritized surface-level features that might signal LLM involvement over the actual quality and substance of the research content. Four primary detection strategies emerged from their explanations:

\textbf{1. Authenticity Detection Through Writing Style \reedit{(N = 18)}:} Participants actively searched for linguistic markers to identify LLM involvement\reedit{, consistent with prior work \cite{Jakesch2023}}. P18 identified \textit{perceived} LLM characteristics in a human-written by stating, \textit{``seems like it was written by a non-english speaker or an AI that cannot structure sentences correctly.''} Similarly, P30 detected LLM patterns in rewritten abstracts, observing that \textit{``A lot of words used in option 1 and 2 hints to use of an LLM rather than human writings.''} Conversely, participants attributed human qualities to LLM-generated content, with P18 praising a generated abstract because it \textit{``feels like something I would easily present or say in real life''} and \textit{``uses words like `We' and `Our' implying ownership.''}

\textbf{2. Structural Coherence as Credibility Signal \reedit{(N = 6)}:} For some, logical flow and organization of the abstract served as primary evaluation criteria when authorship was unknown. P42 valued abstracts that \textit{``first explains the motivation, problem and then gradually flow to the solution and results,''} while P7 emphasized how \textit{``consistent, logically connected and direct sentences...makes the story sounds more logical and trustworthy.''} This focus on narrative structure extended to accessibility, with P26 noting successful abstracts \textit{``strike a balance between clarity and technical depth, making it more accessible, especially for readers who may not have a background in machine learning.''}

\textbf{3. Technical Complexity Paradox \reedit{(N = 12)}:} Technical language created conflicting evaluations. P31 criticized human-writtens as \textit{``Too technical - abstract jumps right into the kernel functions without describing the problem, scope etc. Too technical to no one benefit,''} while P24 complained that generated abstracts contained \textit{``Lots of technical jargons''} making it \textit{``complicated and tough to understand the main goal of the research.''} Yet technical depth also signaled authenticity, with P38 praising abstracts that \textit{``achieve a strong balance between clarity and technical depth.''}

\textbf{4. Natural Language as Human Marker \reedit{(N = 6)}:} Participants associated conversational tone with human authorship. P1 identified human writing by stating that, \textit{``The writing is cohesive and sounds human-like''} and \textit{``well-organized and easy to follow.''} The presence of personal pronouns particularly influenced authenticity judgments, with multiple participants noting the use of ``we'' and ``our'' as indicators of genuine authorship.

These themes reveal that under uncertainty, participants primarily engage in authorship detection rather than quality assessment. When authorship is disclosed, however, a markedly different set of evaluation criteria emerges, focusing on academic merit rather than authenticity signals.
\newline

\noindent\textbf{With Authorship Knowledge (information condition)}
When authorship was disclosed, participants shifted from detection to assessment, evaluating abstracts based on academic merit and functional effectiveness. Their explanations revealed four quality-focused evaluation criteria:

\textbf{1. Evidence-Based Quality Assessment \reedit{(N = 20)}:} When authorship was known, participants focused on quantitative support and specificity. P40 criticized human-writtens noting \textit{``the text does not provide any quantitative elements to support the conclusions of the authors''} and suggested \textit{``I would add quantitative statements to support the claims being made.''} P45 similarly noted for rewritten abstracts that they \textit{``Lack numerical comparison with existing study which would help grabbing attention and show clear advantages,''} while P33 wanted generated abstracts to \textit{``include more specifics, including summarized metrics, related to the results.''}

\textbf{2. Academic Rigor and Publication Standards \reedit{(N = 5)}:} Knowledge of authorship shifted evaluation to meeting scholarly conventions. P15 assessed abstracts against publication criteria, stating \textit{``The abstract is strong in terms of logical flow, readability and length''} and \textit{``With just a few additions and adjustments, this abstract could be accepted for a conference.''} P45 valued methodological transparency, noting abstracts that \textit{``explicitly states that three representative sampling algorithms are used...such detailed information is helpful for reproducing the results which is an essential step to furthering the research.''}

\textbf{3. Functional Communication Effectiveness \reedit{(N = 23)}:} Participants evaluated how well abstracts served their purpose. P34 praised abstracts that \textit{``strike a balance by including enough detail to convey innovation without overwhelming the reader''} and \textit{``avoid redundancy and keep the focus on the novel aspects of the work, making it easier for reviewers and readers to understand why the method is an improvement.''} P23 valued narrative structure, appreciating abstracts with \textit{``the best story-like structure: the author produces the problem, then propose the overall solution, then highlights the contribution.''}

\textbf{4. LLM Enhancement Recognition \reedit{(N = 15)}:} When LLM involvement was disclosed, participants explicitly evaluated its contribution. P15 observed that \textit{``LLM introduced linguistic clarity and cohesiveness,''} while P41 noted LLM-edited abstracts \textit{``provide an organic flow of the ideas of the research with better choice of words.''} However, concerns arose about appropriateness, with P4 noting \textit{``LLM-written text brings too many specific terms from the paper''} and questioning whether certain technical terms \textit{``should probably not be in the abstract.''}
\newline
}

\reedit{{ 
\footnotesize 
\renewcommand{\arraystretch}{1.5} 

\begin{longtable}{p{0.25\textwidth} p{0.70\textwidth}}


\toprule
\textbf{Theme \& Frequency} & \textbf{Key Insight \& Illustrative Quotes} \\
\midrule
\endfirsthead

\multicolumn{2}{c}{{\bfseries \tablename\ \thetable{} -- continued from previous page}} \\
\toprule
\textbf{Theme \& Frequency} & \textbf{Key Insight \& Illustrative Quotes} \\
\midrule
\endhead


\bottomrule
\endfoot

\bottomrule
\caption{Comparison of evaluation strategies under uncertainty (Guess Condition) vs. certainty (Information Condition), resulting in three key transformations in participant reasoning.} 
\label{tab:qual_strategies_comparison} \\
\endlastfoot


\multicolumn{2}{l}{\textbf{\textit{A. Without Authorship Knowledge (Guess): Focus on Authenticity}}} \\ 
\midrule

\textbf{1. Authenticity Detection} \newline (N = 18) & 
Participants operated as "detectives," searching for linguistic errors or markers to identify LLM involvement. \newline
\textit{``Seems like it was written by a non-english speaker or an AI...''} (P18); \textit{``Hints to use of an LLM rather than human writings.''} (P30) \\
\addlinespace

\textbf{2. Structural Coherence} \newline (N = 6) & 
Logical flow served as a proxy for credibility when authorship was unknown. \newline
\textit{``Consistent, logically connected and direct sentences...makes the story sounds more logical and trustworthy.''} (P7) \\
\addlinespace

\textbf{3. Technical Complexity} \newline (N = 12) & 
Technical depth was viewed inconsistently: sometimes as a barrier, other times as a signal of authenticity. \newline
\textit{``Too technical to no one benefit.''} (P31) vs. \textit{``Achieve a strong balance between clarity and technical depth.''} (P38) \\
\addlinespace

\textbf{4. Natural Language} \newline (N = 6) & 
Conversational tone and pronouns were used as heuristics for human authorship. \newline
\textit{``The writing is cohesive and sounds human-like.''} (P1); \textit{``Feels like something I would easily present or say in real life.''} (P18) \\

\midrule
\multicolumn{2}{l}{\textbf{\textit{B. With Authorship Knowledge (Info): Focus on Merit}}} \\ 
\midrule

\textbf{1. Evidence-Based Quality} \newline (N = 20) & 
Shifted focus to the presence of quantitative support and specificity in results. \newline
\textit{``The text does not provide any quantitative elements...''} (P40); \textit{``Include more specifics, including summarized metrics.''} (P33) \\
\addlinespace

\textbf{2. Academic Rigor} \newline (N = 5) & 
Abstracts evaluated against publication standards and reproducibility. \newline
\textit{``With just a few additions... this abstract could be accepted.''} (P15); \textit{``Detailed information is helpful for reproducing the results.''} (P45) \\
\addlinespace

\textbf{3. Functional Effectiveness} \newline (N = 23) & 
Assessed how well the abstract communicated innovation to the intended audience. \newline
\textit{``Avoid redundancy and keep the focus on the novel aspects... making it easier for reviewers.''} (P34) \\
\addlinespace

\textbf{4. LLM Enhancement} \newline (N = 15) & 
Explicit evaluation of the LLM's contribution to clarity vs. potential over-technicality. \newline
\textit{``LLM introduced linguistic clarity...''} (P15); \textit{``LLM-written text brings too many specific terms.''} (P4) \\

\midrule
\multicolumn{2}{l}{\textbf{\textit{C. The Knowledge Effect: Key Transformations}}} \\ 
\midrule

\textbf{1. Framework Shift} & 
\textbf{From Authenticity Detectives $\rightarrow$ Quality Assessors.} \newline
Without knowledge, participants search for "tells"; with knowledge, they evaluate functional adequacy. Identical features receive opposing evaluations based on this lens. \\
\addlinespace

\textbf{2. Technical Reversal} & 
\textbf{From Authenticity Marker $\rightarrow$ Communicative Tool.} \newline
Technical jargon shifts from being a clue about who wrote it (authenticity) to a measure of how well it reads (effectiveness). \\
\addlinespace

\textbf{3. Surface vs. Substance} & 
\textbf{From Stylistic Cues $\rightarrow$ Empirical Evidence.} \newline
Participants abandon surface heuristics (tone, pronouns) in favor of demanding quantitative elements and numerical comparisons. \\

\end{longtable}
}}
\redelete{
\noindent\textbf{The Knowledge Effect: Comparison Between Conditions}
Comparing the two conditions i.e., guess versus information, reveals systematic differences in how participants approach evaluation. We summarize key transformation that describe the impact of authorship knowledge. 

\textbf{1. Evaluation Framework Transformation.} Without authorship knowledge, participants operated as authenticity detectives, searching for linguistic tells and credibility markers. With knowledge, they became quality assessors, evaluating functional adequacy and academic merit. This fundamental shift resulted in opposing evaluations of identical features.

\textbf{2. The Technical Language Reversal.} In the guess condition, technical complexity served dual purposes---both as an authenticity marker (\textit{``too technical to no one benefit''} - P31) and a credibility signal. In the information condition, the same technical language was evaluated purely for communicative effectiveness, with participants like P34 appreciating abstracts that \textit{``strike a balance between technical details and readability''} regardless of authorship.

\textbf{3. From Surface to Substance.} Guess condition participants focused on stylistic elements such as whether writing sounded \textit{``human-like''} (P1) or showed \textit{``ownership''} through pronouns (P18). Information condition participants ignored these surface features, instead demanding \textit{``quantitative elements to support conclusions''} (P40) and \textit{``numerical comparison with existing study''} (P45).

These findings show that knowing who wrote an abstract changes what readers pay attention to. When authorship is unclear, readers focus on detecting LLM involvement rather than evaluating quality, suggesting that disclosure is important for fair assessment.}

\subsection{Orientations Toward LLM-Generated Content and Evaluation Practices (RQ4)}
\label{rq5}

\begin{figure}[t]
  \centering
  \includegraphics[scale=0.5]{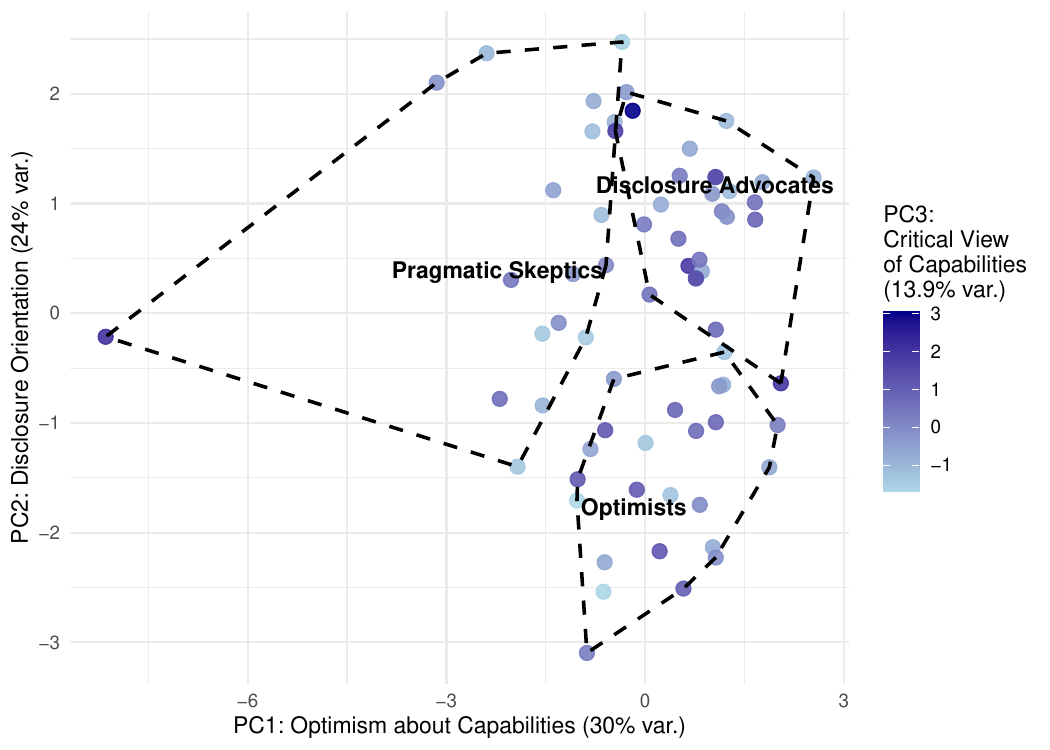}
  \caption{Participants in PCA space (PC1 vs.\ PC2), colored by PC3 (sequential blue scale). Participants fall into three clusters based on their orientation towards LLM-generated content with straight-sided convex-hull enclosures per cluster.}
  \label{fig:pca-clusters}
\end{figure}

Understanding participants' orientation toward LLM-generated content allows us to move beyond average effects and uncover distinct ways in which individuals make sense of and evaluate LLM-authored research summaries. To this end, we derive participant profiles using LLM-related background and attitudes reported in the survey. This analysis is build around eight attitude questions, each rated on a 5-point Likert scale of agreement. The full list of questions can be found in Appendix~\ref{app:full_survey} under the LLM opinion page.

Principal component analysis reveals three main dimensions explaining 67.85\% of the variance in participants' orientations toward LLMs. The first component (30.01\% of variance) represents optimism about LLM capabilities with positive loadings on positive prior experience (0.446), suitability for complex and academic tasks (0.443, 0.448), the viability of summaries (0.341), and the role of LLMs in research (0.455).
The second component (23.99\% of variance) reflects an orientation in favor of LLM use disclosure. It is defined by high loadings on support for disclosure in research and writing (0.606, 0.620), and a negative loading on LLM suitability for academic tasks (-0.333).
The third component (13.86\% of variance) encompasses skepticism towards LLM capabilities with strong negative loading on exaggerated capabilities (-0.863), counterbalanced by moderate loadings on suitability for complex tasks (0.347) and negative loading on prior experience (-0.338). 

Building on these principal components, we clustered participants to identify groups with similar orientations toward LLMs. Results are depicted in Figure~\ref{fig:pca-clusters}. Silhouette analysis combined with k-means revealed three coherent participant profiles (with associated profile means): (1) Disclosure Advocates (39.13\% of participants) reported generally positive orientations including positive prior experience with LLMs (1.15) and rated them as suitable for complex (0.93) and academic (1.15) tasks, with summaries viewed as generally viable (0.70). They stood out for very high support for disclosure in both research and writing (1.59,1.59) and strong endorsement of a role for LLMs in research (1.74). 
(2) Pragmatic Skeptics (26.09\% of participants), expressed more mixed orientations with moderate prior experience (0.72) but concerns about exaggerated capabilities (0.50), limited suitability ratings for complex (–0.50) and academic tasks (+0.11), and low viability judgments (–0.67). They offered modest support for disclosure (0.89 in research, 0.56 in writing) and endorsed LLMs’ role in research (1.00).
(3) Optimists (34.78\% of participants), were highly positive and reported strong prior experience (1.67) and academic suitability (1.46), moderate suitability for complex tasks (0.38), and strong endorsement of LLMs’ role in research (1.75). However, they expressed lower support for unconditional disclosure, disagreeing with the idea that authors should always disclose LLM involvement in research (–0.46) and writing (–0.75).

These clusters were also reflected in participants’ free-text comments.
Disclosure Advocates welcomed AI but stressed the importance of human-AI collaboration calling on researchers to “think, put together their thought and refine them with help of llms \& ultimately own the content” (P31) or noting that “collaborative abstracts between paper authors and LLMs seem consistently better than either the author alone, the LLM alone, or both” (P63).
Pragmatic Skeptics voiced caution, warning that “LLM can never be main source of academic research and it can’t be trusted” (P15) and expressing fears of skill erosion: “just because we can, why should we? [...] I fear we’re going to lose our own ability to do high-quality academic writing if we don’t exercise this muscle” (P55).
Optimists, by contrast, stressed productivity and quality gains, calling LLMs “just tools that could drastically increase a researcher’s productivity, if used properly” (P12) and arguing that “as long as they are factual and relevant they should be used extensively” (P34). Some explicitly opposed mandatory disclosure for writing assistance, cautioning that “forcing disclosure is a bad path [...] [disclosure of] aid 'behind the scenes', I don't think it's needed and could easily become a distraction and a way for LLM bias (in either direction) influencing the research/editorial process in an ineffective way” (P58).

\section{Discussion}
\subsection{Findings in Context}






Our experiment examines how readers engage with research abstracts under varying degrees of LLM involvement.
\redelete{, investigating their ability to detect authorship (RQ1), how actual involvement shapes judgments of quality and trust (RQ2), how beliefs about involvement influence these assessments (RQ3), and what broader orientations toward LLM-generated content emerge in participants’ evaluation practices (RQ4).}
\redelete{We first find that participants could not reliably distinguish between LLM-generated, human-written, or LLM-edited abstracts, tending instead to assume some degree of human involvement (RQ1). At the same time, they showed a baseline suspicion that LLMs were involved across all abstracts. Participants' judgments drew on heuristics such as completeness, clarity, credibility, engagement, and writing conventions, but these cues frequently led to confident misattributions. Clear and fluent LLM text was often praised as distinctly human, while complex or incomplete human writing was mistaken for AI.}
\reedit{We first find that participants do not reliably distinguish between LLM-generated, human-written, or LLM-edited abstracts, tending instead to assume some degree of human involvement (RQ1). At the same time, they show a baseline suspicion that LLMs were involved across all abstracts. 
Participants' judgments draw on heuristics such as completeness, clarity, credibility, engagement, and writing conventions, but these cues prove systematically unreliable. For instance, P31 praised LLM-generated content as written by "a well educated researcher who has a clear understanding of the problem," while P12 criticized a human-written abstract for "unnecessarily complicated phrases (most likely generated by an LLM)." 
}
These findings align with prior work surfacing readers' inability to distinguish LLM and human-written texts in contexts like news articles, story writing, and peer review \citep{clark-etal-2021-thats,Porter2024,Jakesch2023,HADAN2024100095}.
Similar to our observations, \citet{Jakesch2023} describe that their participants used intuitive but often flawed heuristics to detect whether text is LLM-generated. \citet{Porter2024} report on a human preference bias similar to ours where readers tend to judge LLM-generated works as human more often than the reverse.

For the next research question, we find that participants’ evaluations of abstracts reflect both the type of authorship and whether LLM involvement was disclosed. If authorship was disclosed, participants' gave higher quality ratings to human-written and LLM-edited abstracts as compared to LLM-generated ones. \reedit{Quantitatively, LLM-edited abstracts received the highest clarity ratings ($\beta$ = 1.383, $p$ < .001) and were selected by 55\% of participants when authorship was disclosed, compared to 27-28\% for human-written and LLM-generated alternatives.}
\redelete{Human-written abstracts were valued for conciseness and coherent structure, LLM-edited abstracts stood out for clarity and balance and became the most trusted and preferred option, while LLM-generated abstracts were noted for strong organization but criticized for jargon and lack of focus (RQ2).}\reedit{
Participants valued that LLM editing achieved clarity without sacrificing substance. P15 noted that "LLM introduced linguistic clarity and cohesiveness," while P26 praised abstracts that "strike a balance between clarity and technical depth." By contrast, LLM-generated abstracts were criticized for "information overload without focus" and including "too many specific terms from the paper" (P4), while human-written abstracts were valued for coherent structure but sometimes lacked quantitative support.} 
Readers' preferences for text that is human-written or written in collaboration of human and LLM, rather than entirely generated by an LLM, has been reported by past studies \cite{Hitsuwari2023,10.1145/3613904.3642134,10.1145/3711020}.
Across all abstract types, disclosure consistently elevated ratings of trust and quality. Participants expressed neutrality when guessing but reported positive trust once authorship was revealed; even for fully LLM-generated texts. Abstract preference selection reinforced this pattern, with LLM-edited versions strongly preferred, but only when their LLM authorship level is revealed. Together our findings suggest that while abstract type shapes specific strengths and weaknesses, disclosure of LLM involvement may have a more substantial impact on how readers judge trustworthiness and information quality. This finding contrasts the findings of prior work demonstrating decreases in credibility and quality judgments after AI use has been disclosed \cite{cheong2025penalizingtransparencyaidisclosure,Baek2024,Lim_2024}. We discuss this phenomenon further in Section~\ref{sec:discussion_disclosure}.

Moving to the perceived level of LLM involvement, our results show that participants’ beliefs about authorship shaped their evaluations even when inaccurate (RQ3), which is in line with prior oberveations from the literature \citep{10.1145/3613904.3641955,Baek2024,Jakesch2023}. Abstracts believed to be human-written were judged as clearer, more comprehensive, engaging, and concise, reflecting a strong human preference bias. Abstracts thought to be LLM-generated were still credited with clarity and thoroughness, while LLM-edited texts showed no consistent patterns. 
Importantly, this human preference bias is non-directional. Without further study, it is difficult to say whether participants judge abstracts they perceive to human-generated more favorably, or whether they think the abstract is human-written precisely because they perceive it to be of higher quality. \reedit{The qualitative data suggest the latter may be at play: P1 identified human writing because it "sounds human-like" and was "easy to follow," using the very qualities measured in our rating scales to infer authorship.}\citet{Porter2024} argue that readers expect AI-generated writing to be worse, so when they encounter a piece they like, the interpret this preference as signal that the piece is human-written.
Trust, however, did not vary by perceived authorship, reinforcing that disclosure rather than belief is the key determinant of trust. \redelete{Qualitative explanations reveal how uncertainty transforms evaluation: when authorship was unknown, participants acted as authenticity detectives, scanning for linguistic tells or stylistic cues; when authorship was disclosed, they became quality assessors, focusing on evidence, rigor, and communicative effectiveness.}\reedit{Qualitative explanations reveal how uncertainty transforms evaluation. When authorship was unknown, participants acted as authenticity detectives: P18 searched for linguistic markers, stating that an abstract "seems like it was written by a non-english speaker or an AI," while P30 noted that "a lot of words used...hints to use of an LLM rather than human writings." When authorship was disclosed, these detection behaviors subsided. Participants instead became quality assessors, with P40 criticizing abstracts that did "not provide any quantitative elements to support the conclusions" and P15 evaluating whether abstracts met publication standards.} Together these findings suggest that beliefs about authorship guide judgments of specific qualities, but transparency is necessary for trust.

For the last question, our results reveal three distinct participant orientations toward LLM-generated content (RQ4): (1) Disclosure Advocates, who value LLMs but insist on transparency and express preference of human-AI collaboration, (2) Pragmatic Skeptics, who are cautious about exaggerated claims yet still recognize a role for LLMs in research and (3) Optimists, who are highly supportive of LLM integration but express lower agreement with unconditional disclosure requirements.
Tensions between these orientations may explain some of the mixed observations in the literature. For example, \citet{Cardon2025} observe that employees generally accept AI help to make messaged more professional, yet they perceived managers with high AI use to be less sincere. \citet{li-etal-2024-disclosure} report that AI use disclosure decreases reader's wuality ratings on average but introduces much more variability between individuals as some readers are highly critical of AI-assisted text, while others are unaffected. 
\citet{liao2024llmsresearchtoolslarge} conduct a large scale survey and find that over 80\% of academics are already using LLMs, although participants widely acknowledged signficiant risks associated to LLM use including hallucinations and erosion of originality.
Together, these findings reveal that acceptance of LLM-generated content is not monolithic but shaped by distinct orientations which points to the need for nuanced design, policy, and disclosure strategies that account for divergent expectations and trust thresholds.

\subsection{Peer Review, Disclosure, and the Risks of Misidentification}

\reedit{
Our findings highlight a broader inflection point in how the scientific community perceives and evaluates AI-assisted writing. Consistent with concurrent work on reviewer perceptions \citep{HADAN2024100095}, we find that readers struggle to reliably distinguish human-written from LLM-augmented text, and that evaluations of quality hinge more on coherence, structure, and perceived `human touch' than on the true level of AI involvement. Yet our design allows us to isolate the role of disclosure: participants who were told that an abstract involved LLM assistance expressed higher trust, whereas nondisclosure often triggered suspicion, even when the text was written entirely by humans. This result aligns with emerging academic norms that view LLMs as standard tools and increasingly mandate transparency \citep[e.g.][]{liao2024llmsresearchtoolslarge, acm-author-guidelines}.

However, our results also demonstrate how suspicion of AI use can be detrimental. Misidentifying human writing as LLM-generated can carry significant reputational and professional consequences \citep{Gerezgiher2025UMNexpulsion}. Participants in our study sometimes inferred AI involvement from linguistic cues such as fluency, word choice, or structural irregularities. These features can also reflect an author’s language background or disciplinary writing norms. Mere suspicion of AI use can depress judgments of quality and fairness \citep{Kadoma2025,10.1145/3613904.3641955,Baek2024}, raising concerns about indirect discrimination. These risks are amplified when specific words or stylistic patterns become culturally coded as markers of AI authorship, forming `algorithmic folk theories' that readers may use as heuristics for provenance \citep{huang2024trustllmtrustworthinesslargelanguage, liang2024mappingincreasingusellms}.

Taken together, our findings suggest the need for new AI-use norms and policies in the scientific community that prioritize transparency without punitive overreach. By disentangling the effects of disclosure and documenting the risks of misidentification, our study provides empirical grounding for more trustworthy practices in scientific communication.
}

\nadelete{
\subsection{Trust and Disclosure}
\label{sec:discussion_disclosure}

\reedit{
Our results diverge from prior findings of an `AI disclosure penalty', whereby readers decrease their judgments of content they know to have been generated with an LLM. \cite{Baek2024, cheong2025penalizingtransparencyaidisclosure, Lim_2024,liu2022}.
Instead, we find that transparency about LLM involvement increases participants' trust and quality ratings across all abstract types pointing to a more nuanced picture.
A small body of prior work supports this idea. \citet{13-rae-2024} find that disclosure of AI use lowers satisfaction with the author but does not significantly affect content quality ratings. \citet{li-etal-2024-disclosure} similarly report that disclosure reduces average quality ratings but \emph{increases} variability in ratings. 

Importantly, our participant sample consists of ML experts, a group likely to hold more favorable attitudes toward AI. RQ4 also showed that many were “Disclosure Advocates,” explicitly valuing transparency. Context may further shape these perceptions: within scientific communities, disclosure is increasingly viewed as a professional norm \citep{ACL2023_AIWritingAssistancePolicy}. Participants may therefore have interpreted disclosure as a signal of integrity, shifting from “authorship detective” to “quality assessor.”

These findings have important implications for research and practice. First, they caution against assuming a universal disclosure penalty. Transparency may build trust in settings where readers value integrity and expect clear reporting, such as scholarly publishing. Second, disclosure policies may align reader expectations with the actual writing process, reducing misattribution and suspicion.
}

\redelete{
Our results reveal that transparency about LLM involvement increases participants' trust and quality ratings across all abstract types. This observation diverges from much of the prior literature documenting an AI disclosure penalty. \citet{Baek2024} found that AI-generated advertisements were rated as less credible and less creative than human-authored ones. Similarly, \citet{cheong2025penalizingtransparencyaidisclosure} showed that disclosing AI involvement in news writing reliably decreased perceived trustworthiness. \citet{Lim_2024} observed that LLM-generated messages are judged more harshly. In interpersonal communication contexts, recipients trusted emails less when told that AI assisted with the writing process \citep{liu2022}. Together, this body of work suggests that transparency often triggers an AI penalty, whereby readers decrease their judgments of content they know to have been generated with an LLM.

Yet, a small body of work hints towards more nuanced effects of disclosure.
\citet{13-rae-2024} report that while disclosure of AI use reduced satisfaction with the author, it did not significantly lower ratings of the content quality itself. \citet{li-etal-2024-disclosure} find that disclosure decreases mean quality ratings but, at the same time, increases rating variability with some readers becoming more critical and others remaining unaffected or even reassured \citep{li-etal-2024-disclosure}. 
These findings provide a foundation on which to conceptualize our study's results.

First, unlike many of the previous studies, our participant sample consisted of ML experts, a group likely to hold more favorable attitudes toward technology and to see AI assistance favorably in general. RQ4 further showed that a sizable proportion of our participants were `Disclosure Advocates', explicitly valuing transparency. Context may also play a role. Within scientific communities, disclosure is increasingly seen as a professional norm \citep{ACL2023_AIWritingAssistancePolicy}. Participants may therefore have interpreted disclosure as a signal of integrity, enabling them to shift from “authorship detective” mode, i.e. searching for linguistic tells of AI involvement, to “quality assessor” mode, i.e. focusing on quality assessment. Importantly, our data suggest that disclosure did not simply neutralize skepticism but actively elevated trust, allowing participants to credit LLM-edited abstracts for their clarity, comprehensiveness, engaging writing style, and conciseness.

These findings have important implications for research and practice. First, they caution against assuming a universal disclosure penalty. Transparency may build trust in settings where readers value integrity and expect clear reporting, such as scholarly publishing. Second, disclosure policies may align reader expectations with the actual writing process, reducing misattribution and suspicion. Finally, well-designed disclosure interfaces that are clear, consistent, and contextual can help readers focus on the content rather than speculating about authorship.
}

\reedit{\subsection{AI Literacy and the Evolving ``Hermeneutics of Suspicion''}
While our guess condition explicitly prompted participants to identify LLM involvement, we argue that this active evaluation mode is becoming increasingly representative of realistic reading settings. Historically, readers may have approached scientific texts with a default assumption of human authorship. However, as generative AI becomes pervasive in academic workflows \cite{liao2024llmsresearchtoolslarge}, readers are likely developing what Ricoeur termed a ``hermeneutics of suspicion'' \cite{ricoeur1970freud}---a skeptical mode of interpretation where texts are scrutinized for hidden origins rather than taken at face value.

Our qualitative findings (Section 4.3.2) illustrate the cognitive cost of this suspicion: when authorship was uncertain, participants shifted their focus from evaluating scientific merit (e.g., methodological rigor) to ``authenticity detection'' (e.g., scanning for robotic phrasing or perceived ``AI-words'' like \textit{delve}). In a realistic setting, a reader might begin with a presumption of human authorship, but a single ``linguistic tell''---accurate or perceived, such as an ``em-dash'' (as used here)---can trigger a shift into this detective mode, potentially undermining engagement with the research itself.

Therefore, the value of disclosure extends beyond ethical transparency; it serves a functional cognitive role. By explicitly labeling the degree of AI involvement (e.g., ``LLM-edited''), authors can preempt the reader's ``hermeneutics of suspicion,'' effectively granting them permission to bypass the authorship detection stage and return to evaluating the scientific content. However, a key point to note, which may represent a limitation in itself, is that our work studies this involvement at a broad scale. Consequently, we contend that future work is needed to get more granular about LLM involvement. We are at a point where mere disclosure seems to help, but as this becomes the default, we need to further study effective means to communicate this involvement.}
}

\subsection{Trust, Disclosure, and the ``Hermeneutics of Suspicion''}
\label{sec:discussion_disclosure}

\reedit{
Our findings diverge from prior reports of an `AI disclosure penalty' \cite{Baek2024, cheong2025penalizingtransparencyaidisclosure, Lim_2024, liu2022}: transparency about LLM involvement \emph{increased} trust and quality ratings across all abstract types. Prior work shows similarly mixed effects, with disclosure sometimes lowering satisfaction but not content ratings \citep{13-rae-2024} or reducing average ratings while increasing variability \citep{li-etal-2024-disclosure}. This suggests a more nuanced relationship between disclosure and evaluation.

One likely factor is our participant pool of ML experts, many of whom identified as ``Disclosure Advocates'' (RQ4). Within scientific communities, disclosure is increasingly viewed as a professional norm \citep{ACL2023_AIWritingAssistancePolicy}, which may lead readers to interpret transparency as a signal of integrity rather than a reason for discounting quality.

Our qualitative findings offer a further explanation. As generative AI becomes embedded in academic workflows \cite{liao2024llmsresearchtoolslarge}, readers appear to develop what Ricoeur termed a ``hermeneutics of suspicion'' \cite{ricoeur1970freud}: a tendency to scrutinize text for signs of authorship. In our experiment, when authorship was uncertain, participants shifted attention from scientific merit to ``authenticity detection'' (e.g., robotic phrasing or ``AI-words'' like \textit{delve}). Even small stylistic cues could trigger this detective mode and disrupt engagement.

Disclosure thus serves not only ethical transparency but also a cognitive function: it preempts suspicion and allows readers to focus on content rather than origins. These results caution against assuming a universal disclosure penalty and highlight the need for future work on more granular ways of communicating AI involvement as disclosure becomes standard practice.}

\section{Limitations and Future Work}
Our findings should be interpreted in light of several limitations. First, our participant pool was intentionally drawn from ML researchers and practitioners with high levels of expertise and familiarity with LLMs. While this expert population is well-suited for evaluating machine learning abstracts, it limits the generalizability of our findings to other audiences who may hold less favorable attitudes toward AI. Second, our study focuses exclusively on machine learning research abstracts drawn from arXiv, a domain with relatively standardized writing conventions and a shared technical vocabulary. Perceptions of LLM involvement may differ in domains with more narrative styles (e.g., humanities) or high-stakes applications (e.g., medical research). Third, all generated and rewritten abstracts were produced by a single model (Llama 3.1 8B) using fixed prompts, which may introduce model-specific stylistic effects. Results may not fully generalize to outputs from other models. Finally, our survey captures a snapshot of participants’ judgments in a controlled experimental setting. We do not examine how trust evolves over time, whether repeated exposure calibrates perceptions, or whether participants’ evaluations translate into real-world reliance behaviors such as deciding what to read or cite.

Future work can address these limitations in several ways. First, replication with more diverse populations including early-career students, interdisciplinary researchers, and lay readers would shed light on whether disclosure effects generalize beyond expert communities. Second, systematically varying LLMs, prompt styles, and levels of human involvement would allow researchers to disentangle model-specific effects from general patterns in trust calibration. Third, longitudinal and field studies could capture how trust changes with sustained exposure and whether disclosure fatigue or habituation occurs over time. Future experiments should also explore alternative disclosure designs. Finally, connecting perceptions to downstream behavior would provide a more ecologically valid measure of trust. Taken together, these directions would deepen our understanding of how transparency, model design, and reader background interact to shape trust in LLM-assisted scientific writing.

\section{Conclusion}

Our study shows that readers cannot reliably distinguish between human- and LLM-written abstracts, yet their beliefs about LLM involvement strongly shape judgments of trust and quality. We also identify three distinct reader orientations: Disclosure Advocates, Pragmatic Skeptics, and Optimists. 
Our findings complicate the assumption that disclosure always penalizes AI-assisted work. In our study, transparency increased trust even for fully LLM-generated abstracts. Our results underscore the need for nuanced disclosure strategies and for designs that help readers focus on content quality rather than authorship speculation. Future work should extend this research beyond ML experts, explore multiple LLMs and prompt styles, and connect perception to real-world behaviors such as reading and citation choices, ensuring that LLMs strengthen rather than undermine trust in scientific communication.

\bibliographystyle{ACM-Reference-Format}
\bibliography{sample-base}

\appendix

\section{Methodological Details}

\redelete{
\subsection{Participant Demographics}
\label{app:demo}

Our final set of participants includes 69 individuals (53 external and 16 internal) with 16 identifying as Female, 51 as Male, and 2 preferring not to say. 52.17\% of participants are aged between 34 and 55, and 37.68\% are aged between 25 and 34. While 44.93\% identify as native English speakers, the remaining 55.07\% reported fluency in English. 81.16\% of participants hold or were pursuing a doctoral or equivalent degree at the time of the study, with 43.48\% having completed their highest degree over five years ago, 49.28\% within the past five years, and 7.25\% still in progress. Participants represent a mix of professional backgrounds, including 59.42\% industry professionals (e.g., software engineers, data scientists) and 30.43\% academic researchers (e.g., professors, postdoctoral scholars). Participants also reported high levels of familiarity with large language models (LLMs): 26.09\% described themselves as having some knowledge of LLMs and understanding their general principles, 37.68\% identified as having a good understanding how LLMs work including some of the technical or operational details, and 33.33\% attested to having an in-depth understanding on LLMs, including underlying algorithms, training processes, and applications. Only 2.9\% had little or no familiarity with how LLMs function. All participants indicated that they have personally used LLMs in the past.

Taken together, participants' characteristics suggest strong expertise in both reading and evaluating academic machine learning papers and abstracts, as well as familiarity with LLMs and their inner workings. This makes the participant pool well-suited for studying perceptions of LLM-generated content in expert communities.
}

\redelete{
\subsection{Abstract Generation}
\label{sec:abstract_gen}

Abstract Alterations with LLM involvement were generated using Llama 3.1 8B. We employ default parameter of temperature 0.5, and set the maximum generation length to 1,838 tokens which corresponds to the average number of tokens in our data set of original human-written abstracts plus two standard deviations. Note that for we considered punctuations, contractions, and numbers as tokens for this calculations and these are likely to also appear in generated abstracts.
All alterations include the following abstract writing instructions obtained from the ACM website \cite{acm-author-guidelines}:
\begin{itemize}
    \item [] \texttt{The abstract should be 150 to 200 words and consist of short, direct, and complete sentences. It should be informative enough to serve in some cases as a substitute for reading the paper itself. It should state the objectives of the work, summarize the results, and give the main conclusions, but omit future plans and citations.}
\end{itemize}

For LLM-rewritten (or LLM-edited) abstracts, we provide the LLM with the original human-written abstract from the paper with the prompt to rewrite: 
\begin{itemize}
    \item [] \texttt{Abstract guidelines: [ACM INSTRUCTIONS]\\
    Rewrite this Abstract for a scientific paper with title '[PAPER TITLE]' based on the given abstract guidelines. Assume you are the author of the paper and start with 'Abstract:'.\\
    Abstract: [ORIGINAL ABSTRACT]"}
\end{itemize}

For fully LLM-generated abstracts, we extract all paper sections from the pdf stream of the paper, remove the abstract, and then provide the remainder of the paper as input to the model, i.e.
\begin{itemize}
    \item [] \texttt{Abstract guidelines: [ACM INSTRUCTIONS] \\
    Summarize this Paper into an Abstract for a scientific paper with title '[PAPER TITLE]' based on the given abstract guidelines. Assume you are the author of the paper and start with 'Abstract:'. \\
    Paper: [PAPER TEXT]}
\end{itemize}

LLM-involved abstracts are obtained by selecting everything after the marker 'Abstract:' in the generations. As a final step, all types of abstracts undergo very light editing by one of the authors to correct any type-setting errors as described in the main text.

}

\subsection{Textual Characteristics of Abstract Types}
\label{sec:text_analysis}
We use text analysis methods to better understand how the three types of abstracts differ linguistically and structurally. This has a dual purpose: (1) to validate that the linguistic properties of LLM-generated text are consistent with the findings of prior work \citep[e.g.,][]{liang2024mappingincreasingusellms,liang2024monitoringaimodifiedcontentscale, hake2024quality}, and (2) to provide empirical grounding for interpreting participants' responses in the main study. 
Distributions of abstract-level metrics are compared across authorship types using two-sided Mann–Whitney U tests. Complementary figures can be found in Appendix~\ref{sec:app_diff_abstract_figures}.

\begin{figure}[ht]
    \centering
    \begin{subfigure}[t]{0.48\linewidth}
    \centering
    \includegraphics[width=0.95\linewidth]{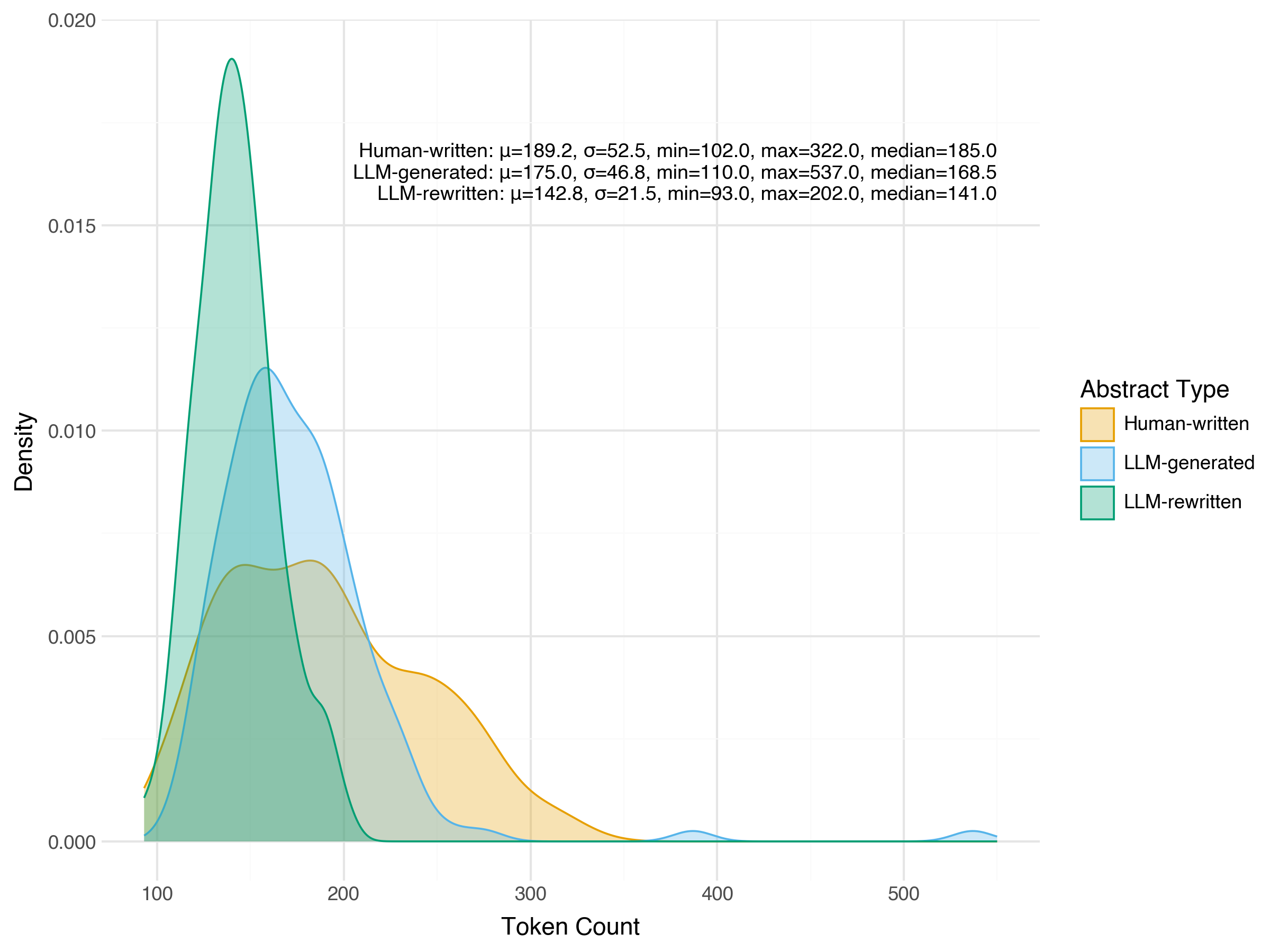}
    \caption{KDE of token counts. Human-written abstracts have more variance in their length, while distributions for abstracts with LLM involvement are more concentrated. LLM-generated abstracts have some outliers with >350 words.}
    \label{fig:token_counts}
    \end{subfigure}
    \hfill
    \begin{subfigure}[t]{0.48\linewidth}
        \centering
        \includegraphics[width=0.95\linewidth]{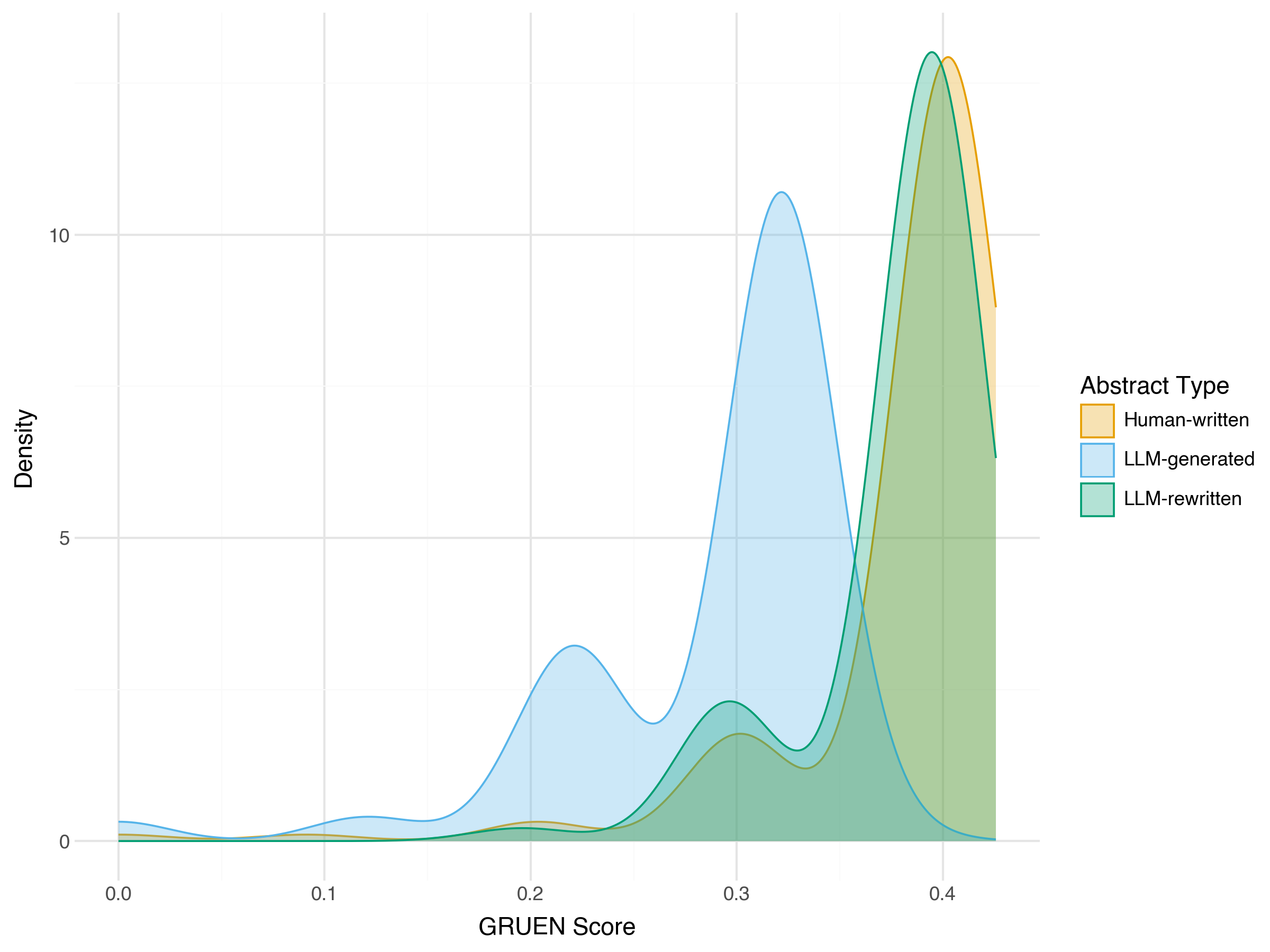}
        \caption{KDE of GRUEN scores \cite{gruen}. Human-written abstracts are of highest linguistic quality with slightly higher scores than LLM-rewritten abstracts. Purely LLM-generated abstracts lag behind in fluency, grammaticality, or coherence.}
        \label{fig:gruen_scores}
    \end{subfigure}
    \begin{subfigure}[t]{\linewidth}
        \centering
        \includegraphics[width=0.95\linewidth]{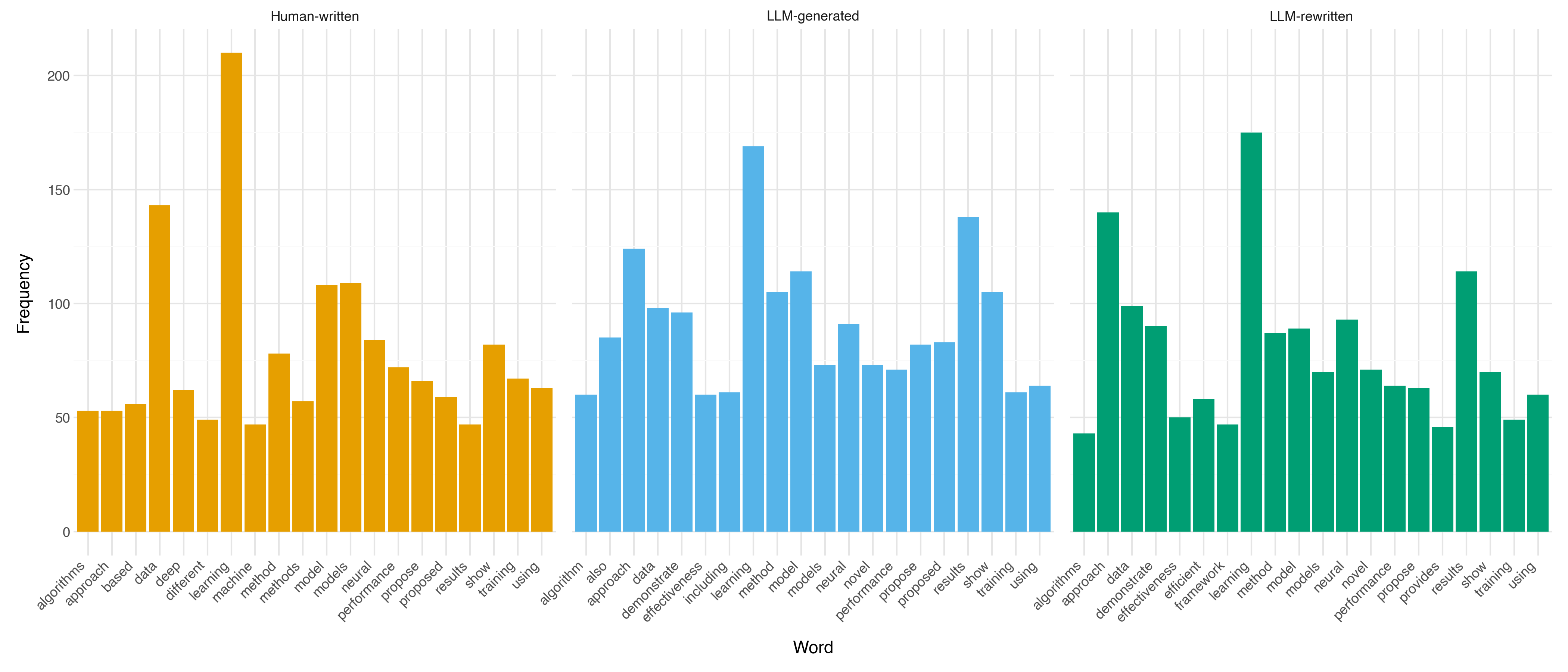}
        \caption{Top 20 most frequent words without stop words. While exact frequencies diverge, there is a large overlap among the most commonly used words (e.g. `learning', `model', `results').}
        \label{fig:top_words}
    \end{subfigure}
    \begin{subfigure}[t]{\linewidth}
        \centering
        \includegraphics[width=0.95\linewidth]{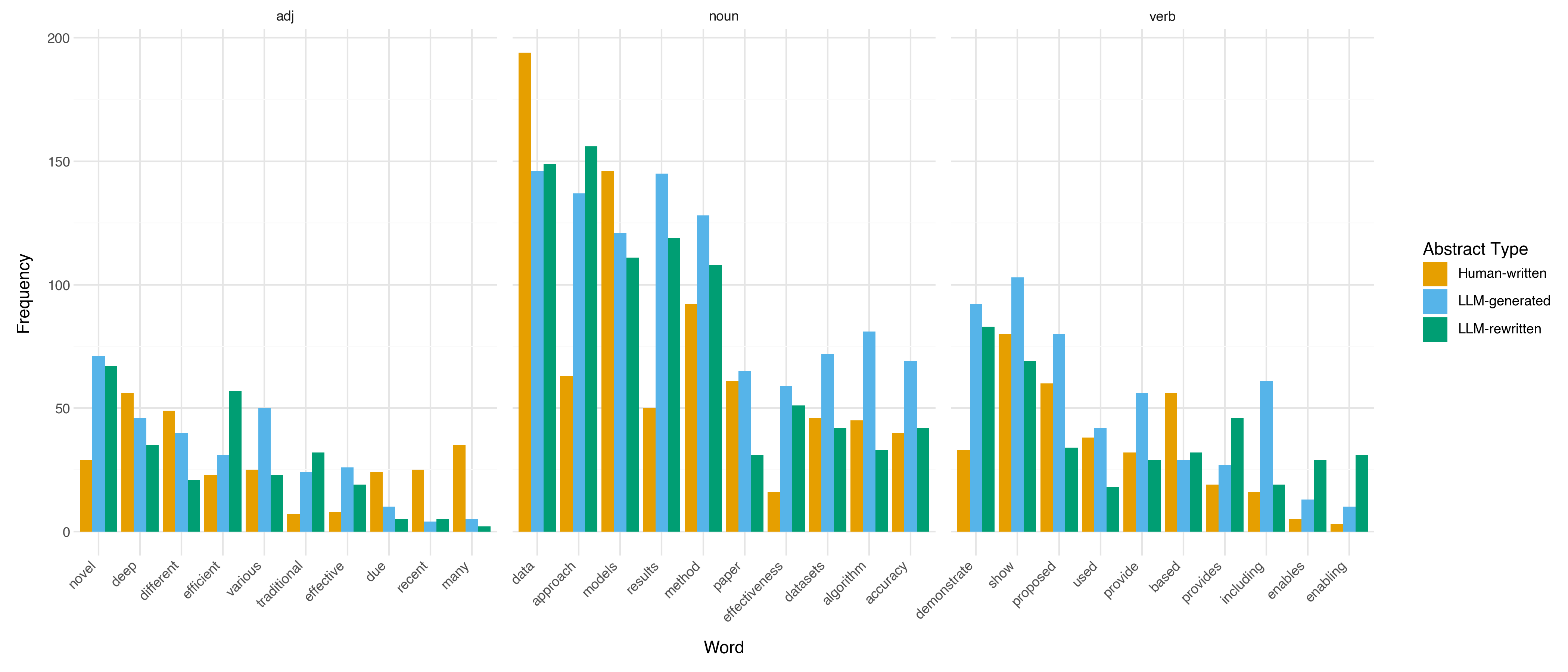}
        \caption{
        Top 10 distinguishing adjectives, nouns and verbs. For each part of speech, we selected the ten words with the highest variance in frequency across abstract types. Human-written abstracts are characterized by words like `due', `recent' and `based', while abstracts with LLM involvement frequently use `novel', `approach', or `demonstrate'.
        }
        \label{fig:distinct_POS}
    \end{subfigure}
    \caption{Linguistic and structural characteristics of abstracts across different generation methods.}
    \label{fig:linguistic_features}
\end{figure}

After tokenizing, human-written (M=189.233 SD=52.478) abstracts in our study are on average longer than LLM-generated (M=175.040, SD=46.760) and LLM-rewritten (M=142.827, SD=21.461) abstracts (one-sided paired t-tests, p = .003 and p < .001). In fact, abstracts with LLM involvement are more consistent in length except for very few outliers of LLM-generated abstracts with >350 tokens (Figure~\ref{fig:token_counts}). 
Both LLM-generated and LLM-edited abstracts remain well below the token limit specified in our prompts, so the cap does not explain the observed length differences. Instead, the finding aligns with prior work reporting that, while LLMs can be verbose in some cases, effective prompting often produces shorter and more concise summaries than those written by humans \citep{hake2024quality, liang2024mappingincreasingusellms}.

We further measure homogeneity of abstracts by computing pairwise cosine similarity of term frequency-inverse documents. Surprisingly, we find that, in our application, the human-written abstracts are most homogeneous (M = 0.101, SD =0.035) as compared to generated (M = 0.098, SD = 0.034) and rewritten (M = 0.081, SD = 0.031) abstracts (p<.001, one-sided paired t-tests) , although absolute differences are small. This suggests that LLMs introduce more lexical variability, particularly when paraphrasing, and points towards strong community conventions that may result in formulaic writing of human-authored abstracts. Higher diversity with LLM was also found in \citet{HADAN2024100095}.

Lastly, we compare the frequency of specific words across abstract types. While the top 20 most frequent words have large overlaps between abstracts with and without LLM involvement (Figure~\ref{fig:top_words}), there are some key words whose use sets abstract types apart. We identify these words by computing the standard deviation of each word’s frequency across abstract types and selecting the words with the highest variability (Figure~\ref{fig:distinct_POS}). We find that human-written abstracts in our sample are particularly likely to use words like `due', `recent', `many', `data', `models', and `based', while abstracts with LLM involvement employ words like `novel', `approach', `effectiveness', `demonstrate', `including', and `enable' more frequently. These observations are in agreement with prior work that has surfaced various lists of specific words associated with LLM use \citep{liang2024monitoringaimodifiedcontentscale, liang2024mappingincreasingusellms}.


\subsection{Figures for Differences between Abstract Types}
\label{sec:app_diff_abstract_figures}

\section{Full Survey Instrument}
\label{app:full_survey}

\emph{Below we provide the complete survey instrument used in our study. The survey was conducted online using a user interface developed for this study. The appendix includes all instructions, questions, and response options as presented to participants. This documentation is provided to support transparency and reproducibility.}

\begin{itemize}
    \item Welcome Page
\end{itemize}
\begin{framed}
\small
\noindent\textbf{Welcome!}

\noindent This research study aims to evaluate how individuals perceive academic abstracts written by humans and Large Language Models (LLMs). Your participation will help us better understand these processes and improve assistive writing tools.

\noindent\textbf{What You Will Do:} You will review a series of titles and abstracts from academic papers. For each abstract, you will answer a series of questions that may ask about quality, clarity, and trustworthiness of the abstract. Some abstracts were written by researchers, while others were generated with the help of Large Language Models (LLMs) using human-written abstracts or entire papers as input. We ask that you approach each abstract independently and provide honest feedback based on your own impressions. There are no right or wrong answers.

\noindent\textbf{Estimated Time Commitment:} The survey should take approximately 60 minutes to complete.

\noindent Your responses will be kept confidential, and no directly identifying information will be collected or shared. If you have any questions about this study, please contact Firstname Lastname at firstnamelastname@example.com.

\noindent By clicking "Next" you confirm that you have read and understood this information and agree to participate in this study.
\end{framed}

\begin{itemize}
    \item Additional Instructions
\end{itemize}
\begin{framed}
\small
\begin{enumerate}
    \item Please do not use the browser back button, click any of the buttons in the top right corner ("Decline task", "Release task", or "Stop and resume later"), or otherwise exit the survey page, as your progress will NOT be saved. Additionally, please do not look up the papers. When you complete the survey, a new one may load; only then should you click the "Release" task button.\\
    \textsquare{} I understand
    \item \emph{PLEASE DO NOT USE AI TOOLS!} Please do not use ChatGPT or any other language models to write your responses or to form impressions about the abstract shown in this study. This research aims to understand peoples' first reactions to content generated by an LLM at different levels. If you use an LLM such as ChatGPT, it will invalidate the experiment. Your genuine, unaided responses are crucial for the accuracy of our findings. Additionally, you do not need to fully understand all the formulas and jargon presented in the study. We are interested in your overall impressions, not in technical expertise. We appreciate your participation and honesty in this research. Thank you!\\
    \textsquare{} I understand
\end{enumerate}
\end{framed}

\begin{itemize}
    \item About You
\end{itemize}
\begin{framed}
\small
\begin{enumerate}[start=3]
    \item  What is your age?\\
    \textsquare{} Under 25\\ \textsquare{} 25-34 \\ \textsquare{} 34-55\\ \textsquare{} 55 or older\\ \textsquare{} Prefer not to say
    \item How do you identify your gender?\\
    \textsquare{} Male \\ \textsquare{} Female\\ \textsquare{} Non-binary\\\textsquare{} Other \\ \textsquare{} Prefer not to say
    \item How would you describe your English proficiency?\\
    \textsquare{} Basic (I can understand and communicate simple sentences)\\ \textsquare{} 
    Fluent (I can communicate effectively in most situations)\\ \textsquare{} Native (English is my first language)
    \item What is your primary (first) language?\\
    \emph{[Free Text]}
    \item Please indicate your highest level of education. \\
    \textsquare{} Secondary school or equivalent\\
    \textsquare{} Post-secondary or associate degree (e.g., technical, vocational, or 2-year college program) \\
    \textsquare{} Bachelor's degree or equivalent (e.g., undergraduate university degree)\\
    \textsquare{} Master's degree or equivalent (e.g., postgraduate degree)\\
    \textsquare{} Doctorate or equivalent (e.g., PhD, professional research degree)
\\
    \textsquare{} Other (describe)
    \item When did you complete your highest level of education?\\
    \textsquare{} I am currently pursuing my degree\\
    \textsquare{} Within the last two years\\
    \textsquare{} 3-5 years ago\\
    \textsquare{} 6-10 years ago\\
    \textsquare{} More than 10 years ago
    \item Have you taken any university-level courses in computer science, data science, or statistics?\\
    \textsquare{} Yes\\
    \textsquare{} No
    \item What best describes your current occupation or primary activity?
    \textsquare{}Student (e.g., university, or postgraduate)\\
    \textsquare{} Academic researcher (e.g., professor, postdoctoral researcher)\\
    \textsquare{} Industry professional (e.g., software engineer, data scientist, business analyst)\\
    \textsquare{} Not currently employed\\
    \textsquare{} Other (describe)
    \item  Have you ever been involved in any of the following roles related to a machine learning or AI conference or journal (e.g., NeurIPS, ICML, ICLR)? Check all that apply.\\
    \textsquare{} Author on a submission\\
    \textsquare{} Attendee\\
    \textsquare{} Reviewer\\
    \textsquare{} Area or Program Chair\\
    \textsquare{} None of the above
    \item  Have you used a Large Language Model (LLM), such as ChatGPT, Claude, or Llama, before? \\
    \textsquare{} Yes \\ \textsquare{} No
    \item How familiar are you with how LLMs function?\\
    \textsquare{} Not at all familiar - I have no knowledge or understanding of how LLMs work\\
    \textsquare{} Slightly familiar - I have a very basic understanding of how LLMs function\\
    \textsquare{} Somewhat familiar - I have some knowledge of LLMs and understand their general principles but not the technical details\\
    \textsquare{} Moderately familiar - I have a good understanding of how LLMs work, including some technical or operational details\\
    \textsquare{} Extremely familiar - I have an in-depth understanding of LLMs, including their underlying algorithms, training processes, and applications
\end{enumerate}
\end{framed}

\begin{itemize}
    \item Task Introduction Page
\end{itemize}
\begin{framed}
\small
\noindent\textbf{You will now begin task 1 out of 3.}

\noindent In this task, you will review the abstract for an academic paper. You will be asked to evaluate its quality, clarity, and trustworthiness. Afterwards, you will be presented with two alternative abstracts for the same paper and asked for your opinions on them. Some of these abstracts are written by researchers, while others have been generated with the help of a Large Language Model (LLM). Please focus solely on the information provided and refrain from looking up the paper or leaving this interface during the study. Your honest, independent impressions are key to the research.

\noindent Remember, there are no right or wrong answers — your insights are what is important.

\noindent When you're ready, click "Next" to begin!
\end{framed}

\begin{itemize}
    \item Large Language Model (LLM) Involvement
\end{itemize}
\begin{framed}
\small
\noindent Please read the following title and abstract for an academic paper carefully before answering the questions below.

\noindent\textbf{[Title] Example Paper Title}\\
\noindent\textbf{[Abstract]} Example Paper Abstract

\begin{enumerate}[start=14]
    \item I recognize the authors of this paper, or I believe I may have read this paper previously.\\
    \textsquare{} Yes\\
    \textsquare{} No
    \item What is the primary research area of the paper?\\
    \emph{[Free Text]}
    \item How familiar are you with this research area? \\
    \textsquare{} Not at all familiar – I have no knowledge or understanding of this research area
    \\ \textsquare{} Slightly familiar – I have heard of this research area but know very little about it
    \\ \textsquare{} Somewhat familiar – I have some knowledge of this research area and understand its basic concepts
    \\ \textsquare{} Moderately familiar – I have a good understanding of this research area, including its common methods and principles
    \\ \textsquare{} Extremely familiar – I have in-depth knowledge of this research area, including advanced concepts, methods, and recent developments
\end{enumerate}
\emph{[Questions (17)-(20) are only asked for participants in the ``guess'' condition. For participants in the ``information'' condition, we reveal the LLM involvement in the authorship of the abstract at the top of the page.]}
\begin{enumerate}[start=17]
    \item How likely do you think it is that a Large Language Model (LLM) was used in any way to write this abstract?\\
    \textsquare{} Not at all likely – I am confident that this abstract was written without any type of LLM assistance
    \\ \textsquare{} Slightly likely – There is a small chance that an LLM was involved in writing this abstract
    \\ \textsquare{} Somewhat likely – It seems possible that an LLM was involved
    \\ \textsquare{} Moderately likely – It seems highly probable that an LLM was involved
    \\ \textsquare{} Extremely likely – I am confident that an LLM was used in some way to write this abstract
    \item I believe, this abstract was:\\
    \textsquare{} Written by the author(s) without LLM assistance
    \\ \textsquare{} Written by the author(s) with assistance from an LLM
    \\ \textsquare{} Written entirely by an LLM
    \item  How confident are you in your assessment of LLM involvement?\\
    \textsquare{} Not at all confident
    \\ \textsquare{} Slightly confident
    \\ \textsquare{} Somewhat confident
    \\ \textsquare{} Moderately confident
    \\ \textsquare{} Absolutely confident
    \item Please explain your reasoning for your answers to the previous questions. Why do you think the abstract was written with or without LLM involvement?\\
    \emph{[Free Text]}
\end{enumerate}
\end{framed}

\newpage
\begin{itemize}
    \item Quality and Trust
\end{itemize}
\begin{framed}
\small
\noindent Now, assume you are reviewing the paper for a major academic conference, but only have access to the abstract at this time. Please fill out the following questionnaire.

\noindent\textbf{[Title] Example Paper Title}\\
\noindent\textbf{[Abstract]} Example Paper Abstract

\begin{enumerate}[start=21]
    \item The abstract ...\\
\end{enumerate}
\begin{tabular}{p{4cm}ccccc}
&Strongly disagree&	Somewhat disagree&	Neutral&	Somewhat agree&	Strongly agree\\
Has appropriate length.& \textsquare{}& \textsquare{}& \textsquare{}& \textsquare{}& \textsquare{}\\
Is free from formatting or typographical errors.& \textsquare{}& \textsquare{}& \textsquare{}& \textsquare{}& \textsquare{}\\
Includes only essential and relevant information.& \textsquare{}& \textsquare{}& \textsquare{}& \textsquare{}& \textsquare{}\\
Uses clear language with sentences that are short, direct, and complete.& \textsquare{}& \textsquare{}& \textsquare{}& \textsquare{}& \textsquare{}\\
Clearly states the objectives of the work.& \textsquare{}& \textsquare{}& \textsquare{}& \textsquare{}& \textsquare{}\\
Provides a clear summary of the results.& \textsquare{}& \textsquare{}& \textsquare{}& \textsquare{}& \textsquare{}\\
Effectively communicates the main conclusions of the research.& \textsquare{}& \textsquare{}& \textsquare{}& \textsquare{}& \textsquare{}\\
Allows readers to grasp the essence of the research without reading the full paper.& \textsquare{}& \textsquare{}& \textsquare{}& \textsquare{}& \textsquare{}\\
Is structured logically and flows smoothly.& \textsquare{}& \textsquare{}& \textsquare{}& \textsquare{}& \textsquare{}\\
Is free from ideological or factual bias.& \textsquare{}& \textsquare{}& \textsquare{}& \textsquare{}& \textsquare{}\\
Presents logical and plausible information.& \textsquare{}& \textsquare{}& \textsquare{}& \textsquare{}& \textsquare{}\\
Is of high quality.& \textsquare{}& \textsquare{}& \textsquare{}& \textsquare{}& \textsquare{}\\
Is trustworthy.& \textsquare{}& \textsquare{}& \textsquare{}& \textsquare{}& \textsquare{}\\
\end{tabular}
\begin{enumerate}[start=22]
    \item Please use your intuition and speculate if necessary. Based on the abstract...\\
\end{enumerate}
\begin{tabular}{p{4cm}ccccc}
&Strongly disagree&	Somewhat disagree&	Neutral&	Somewhat agree&	Strongly agree\\
I am interested in reading the full paper.& \textsquare{}& \textsquare{}& \textsquare{}& \textsquare{}& \textsquare{}\\
I think the research is relevant.& \textsquare{}& \textsquare{}& \textsquare{}& \textsquare{}& \textsquare{}\\
I believe the research’s findings are promising.& \textsquare{}& \textsquare{}& \textsquare{}& \textsquare{}& \textsquare{}\\
I expect the research to be of high quality.& \textsquare{}& \textsquare{}& \textsquare{}& \textsquare{}& \textsquare{}\\
I trust the research.& \textsquare{}& \textsquare{}& \textsquare{}& \textsquare{}& \textsquare{}\\
\end{tabular}
\begin{enumerate}[start=23]
    \item What is something peculiar that stands out to you about this abstract? Please focus only on the writing style of the abstract rather than technical details of the paper it represents.\\
    \emph{[Free Text]}
    \item If you could change one thing about this abstract, what would it be?\\
    \emph{[Free Text]}
\end{enumerate}
\end{framed}

\begin{itemize}
    \item Three Abstracts Page
\end{itemize}
\begin{framed}
\small
\noindent Please carefully re-read the abstract and the two alternative abstracts for the same research paper below and fill out the questionnaire.\\

\begin{tabular}{|c|c|c|}
    \hline
    \textbf{Option 1} & \textbf{Option 2} & \textbf{Option 3} \\
    \hline
    \textbf{[Title] Example Paper Title} & \textbf{[Title] Example Paper Title} & \textbf{[Title] Example Paper Title}\\
    \hline
    \textbf{[Abstract]} Example Paper Abstract 1 & \textbf{[Abstract]} Example Paper Abstract 2 & \textbf{[Abstract]} Example Paper Abstract 3 \\
    \hline
\end{tabular}\\

\emph{[The abstracts are referring to the same paper, and the title remains the same across the options. There is always exactly one human-written abstract, one LLM generated abstract and one human-written and LLM edited abstract. In the `information' condition, we reveal the LLM involvement authorship for each the three abstracts.]}

\begin{enumerate}[start=25]
    \item From your perspective, what are the main differences between these three abstracts? Please clearly write 'Option 1' when referring to the first abstract, etc.\\
    \emph{[Free Text]}
    \item  Which of the three abstracts would you include in the paper?\\
    \textsquare{} Option 1\\
    \textsquare{} Option 2\\
    \textsquare{} Option 3
    \item Please explain your choice.\\
    \emph{[Free Text]}
    \item If given the chance, what would you like to change about the abstract you selected?\\
    \emph{[Free Text]}
\end{enumerate}
\end{framed}

\begin{itemize}
    \item Task Feedback
\end{itemize}
\begin{framed}
\small
\begin{enumerate}[start=29]
    \item This task was...
\end{enumerate}
\begin{tabular}{p{4cm}ccccc}
&Strongly disagree&	Somewhat disagree&	Neutral&	Somewhat agree&	Strongly agree\\
Engaging.& \textsquare{}& \textsquare{}& \textsquare{}& \textsquare{}& \textsquare{}\\
Challenging.& \textsquare{}& \textsquare{}& \textsquare{}& \textsquare{}& \textsquare{}\\
Enjoyable.& \textsquare{}& \textsquare{}& \textsquare{}& \textsquare{}& \textsquare{}\\
\end{tabular}
\begin{enumerate}[start=30]
    \item Do you have any additional comments, questions, or insights about the task?\\
    \emph{[Free Text]}
\end{enumerate}
\end{framed}

\emph{This concludes the first task in the survey. Each participant receives three tasks corresponding to three different papers. For each task, the Task Introduction, LLM Involvement, Quality and Trust, Three Abstracts, and Task Feedback pages are repeated which we omit here for brevity.}

\begin{itemize}
    \item Authorship Reveal Page
\end{itemize}
\begin{framed}
\small
\noindent\emph{This page is only displayed for participants in the `guess' condition.}\\

\noindent You have completed all three tasks! We will now go through the abstracts you selected and reveal whether they were written with help from a Large Language Model (LLM). Please fill out the questionnaire.\\

\begin{tabular}{|p{4.3cm}|p{4.3cm}|p{4.3cm}|}
    \hline
    \textbf{Option 1} & \textbf{Option 2} & \textbf{Option 3} \\
    \hline
    \textbf{[Title] Example Paper Title} & \textbf{[Title] Example Paper Title} & \textbf{[Title] Example Paper Title}\\
    \hline
    \textbf{[Abstract]} Example Paper Abstract 1 & \textbf{[Abstract]} Example Paper Abstract 2 & \textbf{[Abstract]} Example Paper Abstract 3 \\
    \hline
    \textbf{[Source]} This abstract has been written \textbf{\underline{entirely by an LLM.}}&
    \textbf{[Source]} This abstract has been written by the author(s) of the paper \textbf{\underline{without any LLM assistance.}}&
    \textbf{[Source]} This abstract has been written by the author(s) of the paper \textbf{\underline{with assistance from an LLM.}}\\
    \hline
\end{tabular}\\

For task 1, you selected: \textbf{Option \emph{[Selected Option Number]}}

\begin{enumerate}[start=65]
    \item  How much does the source of the abstract surprise you?\\
    \textsquare{} Not at all surprised
    \\ \textsquare{} Slightly surprised
    \\ \textsquare{} Somewhat surprised
    \\ \textsquare{} Moderately surprised
    \\ \textsquare{} Extremely surprised
    \item Would knowing how the abstract was written have influenced which of the three you selected?\\
    \textsquare{} Yes\\
    \textsquare{} No
    \item Please explain.\\
    \emph{[Free Text]}
    \item Given the revealed source, what details in the abstract could hint at how it was written?\\
    \emph{[Free Text]}
\end{enumerate}
\end{framed}

\emph{[The authorship reveal page is repeated for tasks 2 and 3.]}

\begin{itemize}
    \item LLM Opinion Page
\end{itemize}
\begin{framed}
\small
\begin{enumerate}[start=77]
    \item Answer the following: 
\end{enumerate}
\begin{tabular}{p{4cm}ccccc}
&Strongly disagree&	Somewhat disagree&	Neutral&	Somewhat agree&	Strongly agree\\
My prior experience with LLMs has been positive.& \textsquare{}& \textsquare{}& \textsquare{}& \textsquare{}& \textsquare{}\\
The capabilities of LLMs are exaggerated.& \textsquare{}& \textsquare{}& \textsquare{}& \textsquare{}& \textsquare{}\\
LLMs are reliable and trustworthy for producing summaries of complex topics.& \textsquare{}& \textsquare{}& \textsquare{}& \textsquare{}& \textsquare{}\\
LLMs are beneficial and well-suited to assist with academic tasks.& \textsquare{}& \textsquare{}& \textsquare{}& \textsquare{}& \textsquare{}\\
LLM-generated summaries could be a viable alternative to author-written abstracts.& \textsquare{}& \textsquare{}& \textsquare{}& \textsquare{}& \textsquare{}\\
I believe authors should always disclose when they use LLMs during the research process (e.g., for brainstorming, experimentation, or data analysis).& \textsquare{}& \textsquare{}& \textsquare{}& \textsquare{}& \textsquare{}\\
I believe authors should always disclose when they use LLMs in the writing or preparation of their paper (e.g., for drafting, editing, or summarizing content).& \textsquare{}& \textsquare{}& \textsquare{}& \textsquare{}& \textsquare{}\\
I believe LLMs will play a significant role in the future of research communication.	
& \textsquare{}& \textsquare{}& \textsquare{}& \textsquare{}& \textsquare{}\\
\end{tabular}
\begin{enumerate}[start=78]
    \item Do you have any additional comments that you would like to share with us?\\
    \emph{[Free Text]}
\end{enumerate}
\end{framed}

\section{Prompt templates for Qualitative Analysis}
\label{app:qual_analysis_prompts}
\begin{tcolorbox}[
    colback=lightgray!20, 
    colframe=gray, 
    title={\textbf{Exemplar Selection Prompt}}, 
    fonttitle=\bfseries,
    sharp corners=southwest,
    coltitle=black,
    label=prompt:exemplar_selection
]
\scriptsize

\textbf{Task:} The aim of this project is to analyze people's responses to understand their perspectives and reasoning.

Here are responses to the question: ``\texttt{\{question\}}''

\texttt{\{combined\_responses\}}

Please select \texttt{\{num\_exemplars\}} exemplar responses that represent the diversity of perspectives.

\vspace{0.1cm}
\textbf{IMPORTANT SELECTION CRITERIA:}
\begin{enumerate}
    \item \textbf{Analytical Diversity}: Include different types of reasoning and analytical approaches
    \item \textbf{Content Diversity}: Cover different aspects and perspectives mentioned in responses
    \item \textbf{Sentiment Diversity}: Include a balanced mix of positive, negative, and neutral sentiments
    \item \textbf{Response Style Diversity}: Include both detailed analytical responses and more concise/intuitive ones
    \item \textbf{Representativeness}: Choose responses that represent common patterns in the data
\end{enumerate}

\vspace{0.1cm}
\textbf{FORMAT INSTRUCTIONS (CRITICAL):}
\begin{enumerate}
    \item Return ONLY valid JSON that can be parsed by Python's \texttt{json.loads()}
    \item Properly ESCAPE all quotation marks, newlines, and special characters in JSON values
    \item NEVER include raw unescaped quotes, newlines or backslashes in response texts
    \item If a response contains characters that would break JSON, simplify or paraphrase it
\end{enumerate}

\vspace{0.1cm}
\textbf{For each exemplar, provide:}
\begin{enumerate}
    \item The response text (properly escaped for JSON)
    \item Why it was selected (focusing on what type of diversity it contributes)
    \item The type of analysis or reasoning it demonstrates
    \item The sentiment/tone it represents (positive/negative/neutral)
\end{enumerate}

\vspace{0.1cm}
\textbf{Output JSON Structure:}
\begin{verbatim}
{
  "exemplars": [
    {
      "response": "response text",
      "rationale": "why this was selected and what 
                    diversity it contributes",
      "analysis_type": "type of analysis/reasoning 
                        demonstrated",
      "sentiment": "positive/negative/neutral"
    }
  ]
}
\end{verbatim}
\end{tcolorbox}

\begin{tcolorbox}[
    colback=lightgray!20, 
    colframe=gray, 
    title={\textbf{Code Generation from Exemplars Prompt}}, 
    fonttitle=\bfseries,
    sharp corners=southwest,
    coltitle=black,
    label=prompt:code_generation
]
\scriptsize

\textbf{Task:} The aim of this project is to analyze people's responses to understand their perspectives and reasoning.

Here are \texttt{\{num\_exemplars\}} examples for how to generate the codes. For each example, you will see one response with the codes step by step.

These responses are the answer of the question: \texttt{\{question\}}

\vspace{0.1cm}
\textbf{Each generated code has the format like:} \newline
`\textit{quote}' refers to/mentions `\textit{definition of the code}'. Therefore, we got a code: `\textit{code}'.

\vspace{0.1cm}
\textbf{Exemplars:}

\texttt{(1)} \newline
\texttt{Response: \{exemplar['response']\}} \newline
\texttt{Action: Let's code the response.}

\texttt{(2)} \newline
\texttt{Response: \{exemplar['response']\}} \newline
\texttt{Action: Let's code the response.}

\textit{[... for all exemplars ...]}

\vspace{0.1cm}
\textbf{Task:} You need to generate the codes with the format: \newline
`\textit{quote}' refers to/mentions `\textit{definition of the code}'. Therefore, we got a code: `\textit{code}'.

For each exemplar above, please complete the step-by-step coding process.

\end{tcolorbox}

\begin{tcolorbox}[
    colback=lightgray!20, 
    colframe=gray, 
    title={\textbf{Coding All Responses with Exemplars Prompt}}, 
    fonttitle=\bfseries,
    sharp corners=southwest,
    coltitle=black,
    label=prompt:code_all_responses
]
\scriptsize

You are an expert qualitative researcher coding responses about: \texttt{\{question\}}

\textbf{Task:} The aim of this project is to analyze people's responses to understand their perspectives and reasoning.

\vspace{0.1cm}
\textbf{CODING EXAMPLES (showing the approach):}

\texttt{EXAMPLE 1:} \newline
\texttt{Response: \{exemplar['response']\}} \newline
\texttt{Approach: Look for specific observations, themes, concerns, or patterns the participant mentions.}

\texttt{EXAMPLE 2:} \newline
\texttt{Response: \{exemplar['response']\}} \newline
\texttt{Approach: Look for specific observations, themes, concerns, or patterns the participant mentions.}

\textit{[... up to 4 exemplars ...]}

\vspace{0.1cm}
\textbf{CODING INSTRUCTIONS:}
\begin{itemize}
    \item Code each response with specific, descriptive codes that capture what the participant mentioned
    \item Let codes emerge naturally from the data---don't force predefined categories
    \item Be specific about themes, concerns, patterns, or observations mentioned
    \item Create codes that capture the essence of what each participant expressed
    \item New codes can emerge that weren't in the examples above
\end{itemize}

\vspace{0.1cm}
\textbf{RESPONSES TO CODE:}

\texttt{Response 1: \{response text\}} \newline
\texttt{Response 2: \{response text\}} \newline
\textit{[... all responses in batch ...]}

\vspace{0.1cm}
\textbf{FORMAT INSTRUCTIONS (CRITICAL):} \newline
Return ONLY valid JSON with this exact structure:

\begin{verbatim}
{
  "Response 1": {
    "codes": ["code1", "code2", "code3"],
    "definitions": {
      "code1": "Definition of what code1 captures",
      "code2": "Definition of what code2 captures", 
      "code3": "Definition of what code3 captures"
    }
  },
  "Response 2": {
    "codes": ["code4", "code5"],
    "definitions": {
      "code4": "Definition of what code4 captures",
      "code5": "Definition of what code5 captures"
    }
  }
}
\end{verbatim}

For each response above, generate descriptive codes that capture what the participant expressed. Let new codes emerge naturally---don't force predefined categories. Use clear, descriptive code names that capture the essence of observations. Provide clear definitions explaining what each code represents.
\end{tcolorbox}

\begin{tcolorbox}[
    colback=lightgray!20, 
    colframe=gray, 
    title={\textbf{Thematic Analysis Prompt}}, 
    fonttitle=\bfseries,
    sharp corners=southwest,
    coltitle=black,
    label=prompt:thematic_analysis
]
\scriptsize

You are an expert qualitative researcher and thematic analysis specialist with extensive experience in inductive coding and theme development.

You understand that preserving the richness and nuance of qualitative data is paramount. You excel at organizing codes into meaningful themes while maintaining the integrity of each individual code.

Your expertise lies in recognizing that similar-seeming codes often capture distinct nuances that must be preserved.

You don't give mechanical themes, you come up with themes that really communicate the essence of what participants are saying to the question: \texttt{\{question\}}

Since you are doing proper thematic analysis, you also know that codes can go across multiple themes as the goal is to generate high quality themes.

\vspace{0.1cm}
\textbf{Key Principles:}
\begin{itemize}
    \item Preserve richness and nuance of qualitative data
    \item Maintain integrity of each individual code
    \item Recognize that similar codes often capture distinct nuances
    \item Create themes that communicate the essence of participants' responses
    \item Allow codes to appear in multiple themes if appropriate
    \item Focus on generating high quality, meaningful themes (not mechanical categorization)
\end{itemize}

\vspace{0.1cm}
\textbf{Output JSON Structure:}
\begin{verbatim}
{
  "Theme_Name_1": {
    "definition": "Definition of this theme",
    "codes": {
      "CODE_1": "Definition of code 1",
      "CODE_2": "Definition of code 2"
    }
  },
  "Theme_Name_2": {
    "definition": "Definition of this theme",
    "codes": {
      "CODE_3": "Definition of code 3",
      "CODE_4": "Definition of code 4"
    }
  }
}
\end{verbatim}
\end{tcolorbox}

\end{document}